\documentclass[conference]{IEEEtran}
\usepackage[T1]{fontenc}
\usepackage{amsmath,amssymb,booktabs,array,tabularx,graphicx}
\usepackage{microtype,balance}
\usepackage[hidelinks]{hyperref}
\newcommand{\PaperTitle}{MIMO Channel Extrapolation to Unobserved Radiation Modes via Learned Path Refinement}
\author{\IEEEauthorblockN{Shaoyang Sun and Chongwen Huang\textsuperscript{*}}
\IEEEauthorblockA{Zhejiang University, Hangzhou, China\\
3240105985@zju.edu.cn\qquad\textsuperscript{*}Corresponding author: chongwenhuang@zju.edu.cn}}

\usepackage[table]{xcolor}
\definecolor{TableHeader}{RGB}{238,243,245}
\definecolor{TableFocus}{RGB}{228,242,240}
\renewcommand{\arraystretch}{1.09}

\hypersetup{pdftitle={\PaperTitle},pdfauthor={Shaoyang Sun and Chongwen Huang}}
\title{\PaperTitle}
\newcommand{\R}{\mathcal R}
\newcommand{\CN}{\mathcal{CN}}
\newcommand{\jj}{\mathrm{j}}
\newcolumntype{Y}{>{\raggedleft\arraybackslash}X}
\begin{document}
\maketitle
\begin{abstract}
A pattern-reconfigurable multiple-input multiple-output (MIMO) receiver can sample every antenna element and still leave entire radiation modes unobserved. The calibrated patterns weight a common set of propagation paths, which links these missing channels to the measured rows. Our estimator learns to correct the complex gains and spatial frequencies of those paths, so a correction inferred from measured rows updates every unmeasured mode. Each stage combines a damped least-squares proposal with a learned correction scaled by noise power and local array sensitivity. Training on complete simulated channels rewards corrections that reduce prediction error across modes, including those with no measured rows. The learned rule transfers from pursuit-based training estimates to a ridge variable-projection initializer without retraining. In matched simulations, refinement reduces error on wholly unobserved modes as well as across the full channel. Coefficient updates account for most of the gain, and the learned rule further improves on adaptive ridge shrinkage at fixed angles. Individual channel trajectories show that extrapolation can improve even as the residual on the measured rows increases.
\end{abstract}
\begin{IEEEkeywords}
Channel extrapolation, pattern-reconfigurable MIMO, model-based learning, variable projection, multipath estimation.
\end{IEEEkeywords}

\section{Introduction}
Consider a receiver that samples two of four calibrated patterns during one channel snapshot. Every antenna supplies a row of measurements, yet two entire mode matrices remain unobserved. Each propagation path links these matrices. Its angles and complex coefficient are shared across modes, while each calibrated pattern weights it differently.

A path may couple weakly to the sampled patterns and strongly to an unobserved pattern. Noise in its fitted coefficient then propagates disproportionately into the missing channel. We train on complete channels so the estimator can correct this error in shared path coordinates, even when the measured residual gives little indication of it.

Pattern correlation guides training design~\cite{bahceci2017}, and electromagnetic models couple array geometry to radiation patterns~\cite{ying2024em}. PR-Net learns a complex mapping from grouped measurements~\cite{liang2024prnet}. Other schemes acquire a common reference pattern before grouped sampling~\cite{liang2025,liang2026}. TRACE factors a tensor collected over structured pilot slots~\cite{trace2026}. Pattern interpolation also supports pixel-antenna configuration selection~\cite{dinis2026pixel}. We consider one row-wise acquisition with known target calibration. Off-grid pursuit estimates the underlying paths~\cite{mamandipoor2016nomp,zeng2024qnomp}. Beamspace unfolding~\cite{wu2024trimmed} and RC-Flow's channel-space prior and data-consistency projection~\cite{jiang2026rcflow} update channel representations. Our network corrects the paths before evaluating the array responses.

\textbf{Shared-path refinement.} We combine a joint analytic step with bounded corrections to each complex coefficient and its two angular coordinates. The coefficient update combines the fitted path phase with the phase of its backprojected residual. Its scale adapts to the energy with which that path is observed. The same 774-parameter network serves every path and stage. Trained on pursuit estimates, it transfers to a separately tuned variable-projection (VarPro) fit.

\textbf{Extrapolation to unobserved modes.} At 0~dB reference SNR, sampling two modes and reconstructing all four, it reduces mean full-channel error by 11.73\% relative to ridge VarPro. Learning only coefficient corrections recovers 89.6\% of that reduction. Scored on the two wholly unobserved modes alone, refinement lowers mean NMSE by 11.37\% relative to VarPro and by 3.49\% relative to tuned adaptive ridge.

\begin{figure*}[t]\centering
\includegraphics[width=\textwidth]{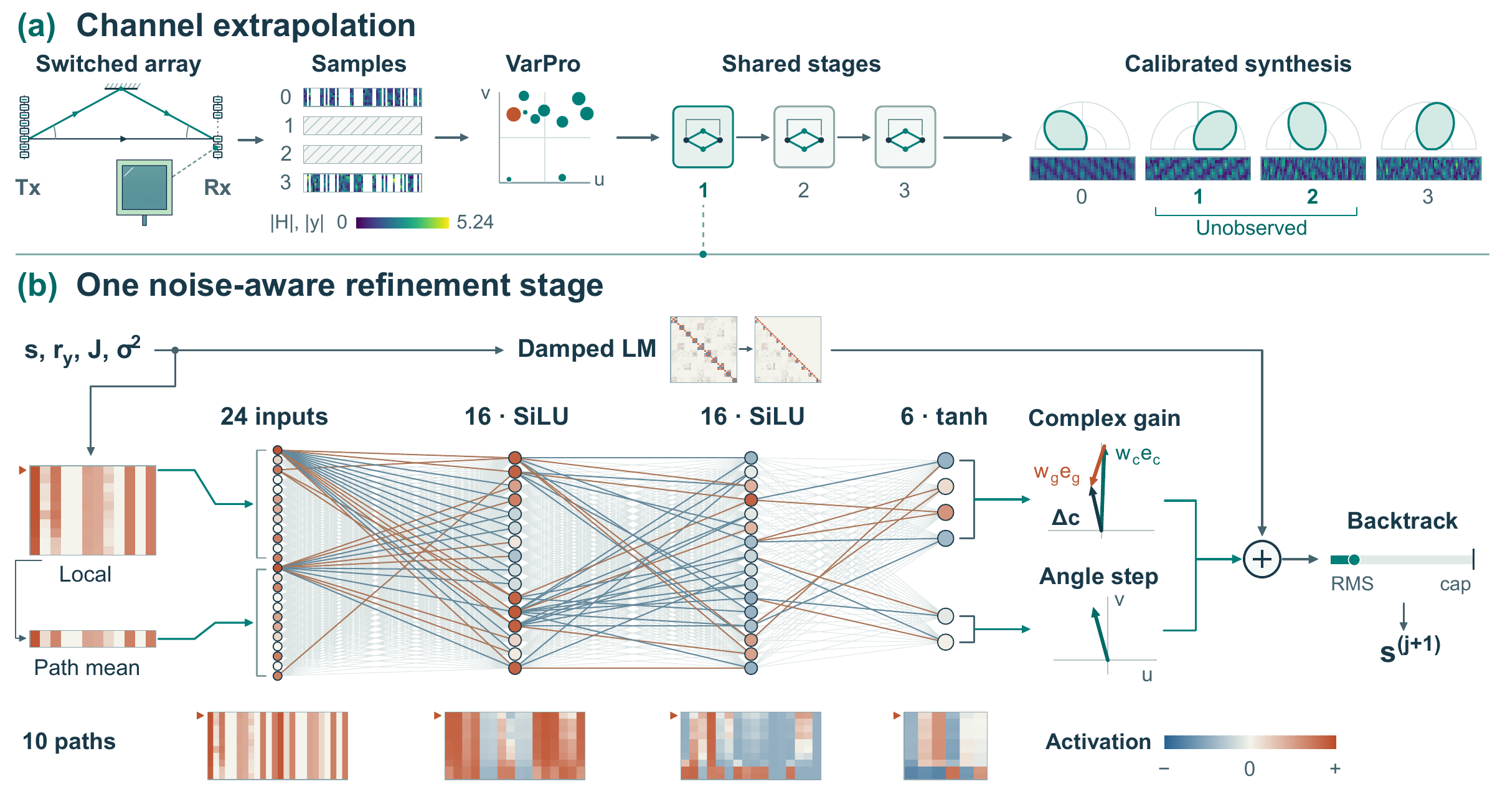}
\caption{Channel extrapolation through shared path refinement. (a) VarPro fits ten paths from modes 0 and 3. The inset sketches a receive element, and hatching marks unsampled modes. After three shared stages, each calibrated pattern (polar plot) renders its channel on the common magnitude scale. (b) Stage one is expanded for the orange path. Node colors encode activations. Highlighted edges carry each neuron's two largest incoming $|W_{ij}h_j|$ terms. The bottom maps show all ten paths, with negative values in blue and positive values in orange. Four outputs weight the two complex responses. The arrows include the common scale in~\eqref{eq:dc}. Two outputs shift the spatial frequencies. The learned correction and the damped LM step are added and tested against the channel-displacement cap.}
\label{fig:framework}\end{figure*}

\section{Acquisition and Extrapolation Target}
Let $c_\ell$ be the complex coefficient of path $\ell$ and $(u_\ell,v_\ell)$ its receive and transmit spatial frequencies. The channel for receive mode $k$ is the superposition
\begin{equation}
[\R(s)]_{krt}=\sum_{\ell=1}^L c_\ell\,g_k(u_\ell)\,e^{\jj\pi(ru_\ell-tv_\ell)},\label{eq:render}
\end{equation}
where $g_k$ is the calibrated field gain, $s=\{c_\ell,u_\ell,v_\ell\}_{\ell=1}^L$, and the receive and transmit indices $r,t$ start at zero. The arrays have unnormalized steering vectors. At row $r$ the receiver selects a single mode $k_r$ for all transmit elements. With $S$ denoting the corresponding row-selection operator, the noisy measurements satisfy
\begin{equation}
 y=S\operatorname{vec}(H)+n,\qquad n\sim\CN(0,\sigma^2 I_m).\label{eq:measurement}
\end{equation}
The receiver knows $y$, the row assignments, the descriptors of the acquired and target modes, and $\sigma^2$. Matching descriptors determines $k_r$ in the target ordering.

Both acquisition masks measure $m=512$ complex entries of a $K\times N_r\times N_t=4\times64\times8$ channel. The four-mode mask assigns 16 rows to each mode. The two-mode mask assigns 32 rows each to modes 0 and 3. Modes 1 and 2 then have known calibration and no measured rows (Fig.~\ref{fig:framework}). Training includes all four modes. Both masks measure the same number of channel entries. Pilot-symbol cost is implementation dependent.

\section{Refinement in Shared Path Coordinates}
\subsection{How measurement noise reaches an unobserved mode}
For a fixed geometry, let $B$ contain the full-channel atoms in~\eqref{eq:render} and $A=SB$ their observed entries. Let $B_U$ restrict $B$ to the modes with no measurements. Even a single path illustrates the transfer of noise. For matched observations $y=ac+n$ with $a\ne0$, unobserved atom $b_U$, and least-squares estimate $\hat c=a^Hy/\|a\|^2$, the expected squared error in the unobserved channel is
\begin{equation}
\mathbb E_n\|b_U(\hat c-c)\|^2=\sigma^2\frac{\|b_U\|^2}{\|a\|^2}.\label{eq:transfer}
\end{equation}
The denominator measures what the receiver learns about this path from its samples. The numerator measures how strongly an error in the coefficient appears on the target modes.

For several paths, ridge regression minimizes $\|Az-y\|^2/m+\lambda_P\|z\|^2$ and gives $\hat c_P=Qy$, with $Q=(A^HA+m\lambda_PI)^{-1}A^H$. For $y=Ac+n$ and a fixed real scale $t$, its target-mode risk after rescaling is
\begin{align}
\mathbb E_n\|B_U(tQy-c)\|^2
&=\|B_U(tQA-I)c\|^2\nonumber\\[-2pt]
&\quad+t^2\sigma^2\|B_UQ\|_F^2.\label{eq:risk}
\end{align}
Reducing $t$ from unity toward zero lowers the variance term. Its effect on squared bias depends on the coefficients and target patterns. This motivates a path-dependent correction whose size depends on measurement sensitivity and whose training objective reflects errors across modes. Equations~\eqref{eq:transfer} and~\eqref{eq:risk} hold the geometry fixed. The estimator also refines uncertain angles.

\subsection{A joint proposal from the measured array}
During training, Newtonized orthogonal matching pursuit (NOMP) detects ten paths on a $96\times48$ grid and refines them off-grid. At inference, a $128\times64$ greedy search initializes joint ridge VarPro~\cite{golub1973varpro,golub2003varpro}. At each trial geometry, VarPro solves the linear coefficient problem, using the development-selected penalty $\lambda_P=\max(10^{-10},0.05\sigma^2)$. The refinement receives this state and the original observations. VarPro uses analytic pattern gains, and the refinement evaluates the supplied gain and slope samples by cubic Hermite interpolation at 129 knots.

At each stage, $r_y=y-S\operatorname{vec}(\R(s))$ is the measured residual. Write $\phi_\ell$ for a vectorized path atom. The observed Jacobian block for the four real coordinates of that path is
\begin{equation}
 J_\ell=S[\phi_\ell,\jj\phi_\ell,c_\ell\partial_u\phi_\ell,c_\ell\partial_v\phi_\ell].\label{eq:jacobian}
\end{equation}
The receive derivative includes the array phase and pattern slope. Let $D_c$ contain the floored column norms on its diagonal and set $\bar J=JD_c^{-1}$. The Levenberg--Marquardt (LM) proposal uses the joint curvature of all measured paths,
\begin{equation}
 \delta_{\rm LM}=D_c^{-1}[\Re(\bar J^H\bar J)+\mu I]^{-1}\Re(\bar J^H r_y),\label{eq:lm}
\end{equation}
where $\mu=0.01(1+\sigma^2/P_0)$. The solve uses a real $4L\times4L$ Cholesky factorization. $P_0$ is the mean per-entry power of the omnidirectional training channels.

\subsection{Learning how far each path should move}
The learned branch in Fig.~\ref{fig:framework}(b) uses the strength of each fitted path, its agreement with the residual, and its sensitivity to perturbations in the presence of other paths. A shared multilayer perceptron (MLP) takes twelve local features and their twelve means over the path set. These include coherence between observed atoms, stage index and noise level. Two SiLU hidden layers of width 16 and a final $\tanh$ yield six real outputs $q_1,\ldots,q_6$. The supplement gives the complete feature vector.

For a path with observed atom $a_\ell=S\phi_\ell$, write $b_\ell=a_\ell^Hr_y$, $d_\ell$ for the floored atom energy, and $\nu=\sigma^2+10^{-6}P_0$. Suppressing $\ell$, define
\begin{equation}
 s_c=\sqrt{\nu/d},\quad e_c=\frac{c}{\sqrt{|c|^2+s_c^2}},\quad
 e_g=\frac{b}{\sqrt{|b|^2+\nu d}}.\label{eq:bases}
\end{equation}
The coefficient scale $s_c$ follows the observed-atom dependence in~\eqref{eq:transfer}. The two responses retain the phases of the current coefficient and the residual projected onto its atom. The MLP forms the complex weights $w_c=q_1+\jj q_2$ and $w_g=q_3+\jj q_4$ in Fig.~\ref{fig:framework}(b), giving
\begin{equation}
 \Delta c=\frac{s_c}{\sqrt2}\left[(q_1+\jj q_2)e_c+(q_3+\jj q_4)e_g\right].\label{eq:dc}
\end{equation}
For $w\in\{u,v\}$, let $d_w$ be the squared norm of the corresponding observed Jacobian column. With $q_u=q_5$, $q_v=q_6$, the angular correction is
\begin{equation}
 \Delta w=q_w\sqrt{\frac{\nu}{d_w+\nu/h_w^2}},\quad h_u=\sqrt3/95,\quad h_v=2/48.\label{eq:angle}
\end{equation}
Consequently $|\Delta c|\leq2s_c$ and $|\Delta w|\leq h_w$. The network can adapt a weakly observed path more strongly while respecting explicit bounds. Reordering the paths reorders their updates. A common phase rotation of $y$ and $c$ rotates the coefficient correction and leaves the angular correction invariant.

\subsection{Accepting the step in channel space}
The algorithm adds the learned increments to $\delta_{\rm LM}$. For step scales $1,1/2,\ldots,1/256$, it clips each candidate $u$ to $[-\sqrt3/2,\sqrt3/2]$ and wraps $v$ modulo two. It takes the first candidate $s^+$ for which
\begin{equation}
 \frac{\|\R(s^+)-\R(s)\|_F}{\sqrt{2048}}
 \leq\frac{\|r_y\|_2}{2\sqrt{512}}+10^{-12}.\label{eq:cap}
\end{equation}
If every candidate exceeds this limit, the current state is retained. The test bounds how much the rendered channel moves. An accepted step may increase the measured residual, which allows training to reward a better extrapolation despite a looser fit to noisy entries. In the first stage shown in Fig.~\ref{fig:framework}(b), the residual response partly cancels the coefficient response in the complex plane. Their resultant offsets part of the analytic gain step. The combined proposal uses 16.4\% of the channel-displacement allowance and is accepted at unit scale. The final state after three weight-shared stages gives $\widehat H=\R(s^{(3)})$. Supervision minimizes full-channel normalized mean-square error (NMSE), differentiating through the accepted state with the discrete scale selection held fixed.

\section{Experiments}
\subsection{Matched observations and comparisons}
We draw receive and transmit angles independently and uniformly on $[-60^\circ,60^\circ]$ and $[-90^\circ,90^\circ]$, respectively. The ten circular complex Gaussian coefficients are jointly normalized to unit energy. Training uses 5000 channels, each sampled under both balanced masks at 0, 10 and 20~dB reference SNR, $\rho_{\rm ref}=10\log_{10}(P_0/\sigma^2)$, with $P_0=0.9997065$. Row assignments are permuted per view. Each of five prespecified seeds uses one pass through the 30\,000 views (3750 Adam steps, batch size eight, learning rate $10^{-3}$). Final weights are evaluated throughout. A separate 500-channel development set supported model and baseline selection.

Two ablations identify what the network learns. \emph{Coefficient-only} keeps the analytic angle step and learns only $\Delta c$. \emph{Noise/stage} retains every output but conditions the MLP only on noise level and stage. We also compare with a deterministic, fixed-angle \emph{adaptive ridge} estimator. Starting from $\hat c_P$, it solves the coupled coefficient system three times, adding a diagonal penalty $3\sigma^2/(|c_\ell^{(j)}|^2+\sigma^2/\|a_\ell\|^2)$ to $A^HA$ at iteration $j$. This rule minimized pooled error among 24 candidates on 32 additional development channels.

The main test uses 1152 independent channels. Estimators see the same noisy entries within each condition, and both masks share the underlying channel and full-array noise draw at a given SNR. We average $\|\widehat H-H\|_F^2/\|H\|_F^2$ over fitted models within channel, then across channels in linear units. Percentage reductions use the named comparator's mean as denominator. The six prespecified comparisons at 0~dB test full refinement against VarPro and the two ablations under both masks. Their two-sided 99.1667\% percentile intervals use 100\,000 common paired-channel bootstrap resamples (95\% Bonferroni familywise level). Other intervals are descriptive paired 95\% intervals conditional on the fitted weights.

\begin{figure*}[t]\centering
\includegraphics[width=\textwidth]{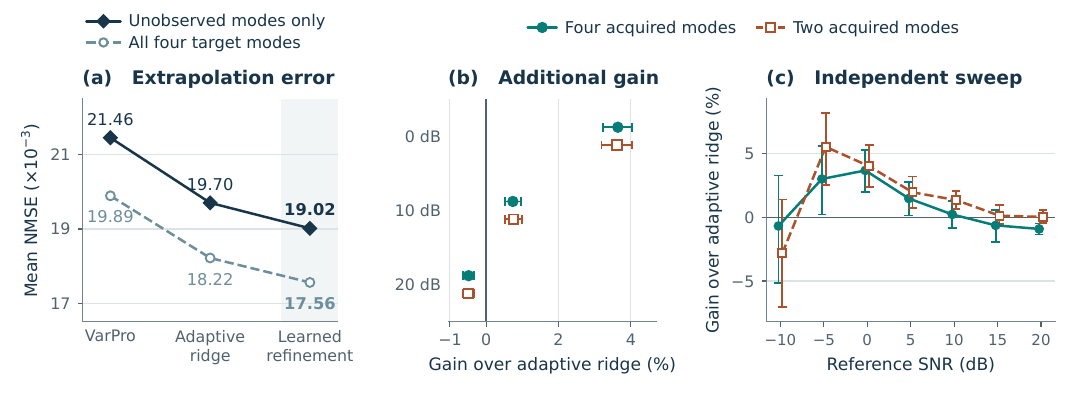}
\caption{Extrapolation accuracy and gain beyond adaptive ridge. (a) Two modes sampled, 1152 channels, 0~dB reference SNR. Each curve normalizes squared error by the energy of its own target, either the two unobserved modes or all four modes. (b), (c) Reduction in mean full-channel NMSE relative to adaptive ridge on the confirmation set and an independent 64-channel sweep. Bars are descriptive 95\% paired-channel intervals.}
\label{fig:evidence}\end{figure*}

\begin{figure}[t]\centering
\includegraphics[width=\columnwidth]{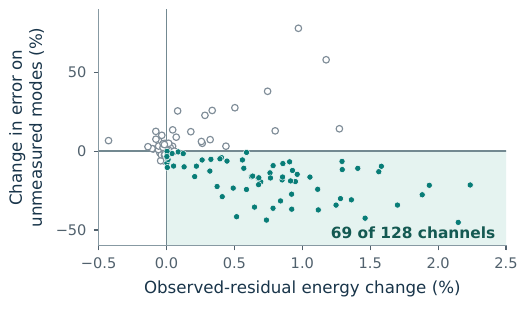}
\caption{Fitting versus extrapolation under two-mode sampling at 0~dB. Each point compares the Hermite-rendered initial state with the final estimate for one of the first 128 confirmation channels, averaging errors across fits at each endpoint. In the shaded quadrant, the residual energy on sampled rows rises while error on modes 1 and 2 falls.}
\label{fig:residual}\end{figure}

\subsection{Accuracy on modes with no measured rows}
\begin{table}[t]
\caption{Mean full-channel NMSE on 1152 confirmation channels ($\times 10^{-3}$). Columns give reference SNR in dB. Each learned result averages five fits. Bold identifies the lowest mean in each column.}
\label{tab:absolute}\centering\footnotesize\setlength{\tabcolsep}{1.7pt}\renewcommand{\arraystretch}{1.18}
\begin{tabularx}{\columnwidth}{@{}lYYYYYY@{}}\toprule
 & \multicolumn{3}{c}{Four modes} & \multicolumn{3}{c}{Two modes} \\
\cmidrule(lr){2-4}\cmidrule(l){5-7}
\rowcolor{TableHeader}Estimator & 0 & 10 & 20 & 0 & 10 & 20 \\\midrule
Direct MLP & 765.6 & 683.0 & 674.2 & 752.0 & 669.8 & 661.2 \\
Pop. LMMSE & 275.1 & 49.4 & 6.99 & 290.7 & 59.8 & 9.21 \\\addlinespace[2pt]
VarPro & 17.60 & 1.425 & 0.1361 & 19.89 & 1.540 & 0.1483 \\
Adaptive ridge & 16.17 & 1.401 & \textbf{0.1358} & 18.22 & 1.517 & \textbf{0.1478} \\
Noise/stage & 17.18 & 1.422 & 0.1362 & 19.39 & 1.537 & 0.1483 \\
Coefficient-only & 15.79 & 1.397 & 0.1367 & 17.80 & 1.512 & 0.1489 \\
\rowcolor{TableFocus}Learned refinement & \textbf{15.58} & \textbf{1.390} & 0.1365 & \textbf{17.56} & \textbf{1.505} & 0.1485 \\
\bottomrule\end{tabularx}\end{table}

At 0~dB reference SNR the median received SNR was 4.78~dB with four modes sampled and 4.83~dB with two, using noiseless power on the selected entries. With two sampled modes, full refinement reduced mean full-channel NMSE from 0.01989 to 0.01756, a reduction of \textbf{11.73\%} with adjusted interval [10.59, 12.84]\%. When all four modes were sampled, the reduction was 11.48\% [10.33, 12.62]\% (Table~\ref{tab:absolute}).

How much of this gain reaches the modes that supplied no data? We replayed the same 1152 two-mode acquisitions and scored only $U=\{1,2\}$, normalizing each channel's squared error by $\|H_U\|_F^2$. Mean NMSE on these wholly unobserved modes fell from 0.02146 with VarPro and 0.01970 with adaptive ridge to \textbf{0.01902} with learned refinement (Fig.~\ref{fig:evidence}(a)). The reductions were \textbf{11.37\%} [10.46, 12.29]\% and \textbf{3.49\%} [3.01, 3.96]\%, respectively. These post-hoc intervals use 10\,000 paired-channel resamples. With the full-channel energy denominator restored, the two missing modes account for 52.6\% of the absolute full-channel NMSE reduction relative to VarPro.

For comparison, population linear minimum mean-square error (LMMSE) estimation gives a mean full-channel NMSE of 0.291 in the two-mode case. A PR-Net-style complex MLP adapted to these masks takes zero-filled observations, a mask and the noise level. Its three complex hidden layers of width 512 train on the same views for 3752 optimizer steps. Adding an LMMSE front end and skip connection brings it close to LMMSE alone. The direct and linear estimators thus leave a substantial gap to the path estimators under the matched sparse channel law.

\subsection{What the learned update contributes}
Learning $\Delta c$ alone recovered 89.4\% and 89.6\% of the full improvement over VarPro for four- and two-mode sampling. Learning $\Delta u,\Delta v$ as well lowered coefficient-only error by a further 1.35\% and 1.37\%. Restricting the MLP to the noise level and stage index removes most of its advantage. Restoring the residual and path features reduced that ablation's error by 9.31\% and 9.45\%. These comparisons locate most of the benefit in \emph{path-specific coefficient correction}.

Adaptive ridge provides a more demanding test of that interpretation. On the same confirmation channels, its coupled shrinkage rule reduced VarPro error by 8.14\% and 8.42\% at 0~dB. Learned refinement lowered the remaining mean error by another \textbf{3.64\%} [3.23, 4.04]\% and \textbf{3.62\%} [3.19, 4.04]\%, respectively (Fig.~\ref{fig:evidence}). All five fits gave this ordering. A separate controller trained with $e_g$ in both terms of~\eqref{eq:dc} retained the same output count and uniform bound. The mixed responses gained a further 1.03 and 1.34 percentage points over this control relative to VarPro.

The first 128 test channels reveal how refinement changes the fit. The learned gain increment had a negative projection onto the current coefficient in 99.93\% of path updates with four sampled modes and 99.76\% with two. On 69 of these 128 channels under two-mode sampling, the fit-averaged endpoints in Fig.~\ref{fig:residual} show both higher measured-residual energy and lower error on the wholly unobserved modes. The acceptance test in~\eqref{eq:cap} permits these corrections.

\subsection{Noise, model mismatch and computation}
An independent 64-channel sweep tests the frozen models between and below training noise levels. At $-5$~dB reference SNR (median received SNR $-0.23$~dB), full refinement reduced VarPro error by 18.70\% and 22.22\% under four- and two-mode sampling. It also improved on adaptive ridge by 3.02\% and 5.54\%. At $-10$~dB reference SNR (about $-5.2$~dB received), both intervals against adaptive ridge include zero. At 20~dB on the larger confirmation set, adaptive ridge has about 0.5\% lower mean error. The clearest advantage over deterministic shrinkage lies between these extremes.

\begin{table}[t]
\caption{Model mismatch at 0~dB reference SNR on 128 additional channels. NMSE is for learned refinement ($\times10^{-3}$). Gain is its reduction relative to ten-slot VarPro. Calibration labels give gain/phase standard deviations.}
\label{tab:stress}\centering\footnotesize\setlength{\tabcolsep}{1.8pt}\renewcommand{\arraystretch}{1.18}
\begin{tabularx}{\columnwidth}{@{}lYYYY@{}}\toprule
 & \multicolumn{2}{c}{Four modes} & \multicolumn{2}{c}{Two modes} \\
\cmidrule(lr){2-3}\cmidrule(l){4-5}
\rowcolor{TableHeader}True channel & NMSE & Gain (\%) & NMSE & Gain (\%) \\\midrule
6 paths & 12.31 & 55.08 & 14.38 & 51.56 \\
8 paths & 13.09 & 40.56 & 15.57 & 36.79 \\
\rowcolor{TableHeader}10 paths (nominal) & 15.65 & 9.67 & 17.62 & 10.09 \\
12 paths & 33.62 & 0.65 & 36.84 & $-0.09$ \\
14 paths & 66.32 & $-0.003$ & 70.42 & $-0.25$ \\\addlinespace[2pt]
0.5~dB / $5^\circ$ SD & 24.99 & 6.21 & 31.66 & 4.89 \\
1~dB / $10^\circ$ SD & 52.75 & 3.51 & 73.25 & 1.60 \\\addlinespace[2pt]
Half power & 33.12 & 16.68 & 38.21 & 14.69 \\
Double power & 7.26 & 6.60 & 8.12 & 6.28 \\
\bottomrule\end{tabularx}\end{table}

We also applied the original ten-slot estimator to 128 additional channels with changes in true path count, pattern calibration or signal power (Table~\ref{tab:stress}). Refinement substantially reduces the penalty for excess slots. With six true paths and two modes sampled it lowers ten-slot VarPro error by 51.56\% at 0~dB. Giving VarPro the true count yields lower error still. With twelve or fourteen true paths, the gain over ten-slot VarPro stays below 1\% in magnitude at this SNR. Under joint per-mode gain and phase errors with standard deviations of 1~dB and $10^\circ$, the two-mode NMSE rises to 0.07325. The supplement details the 54 stress conditions and additional clustered-delay-line tests.

\balance
With a fresh path fit for each call, one-thread CPU inference took 540~ms for VarPro, 542~ms with adaptive ridge and 610~ms with learned refinement (medians on twelve inputs, three timed calls after warmup). A separate two-thread profile attributed 393.4~ms on average to initialization and 0.46~ms to the MLP across all stages. The network adds few parameters, while path fitting and evaluating candidate states account for most of the runtime.

\section{Conclusion}
Learning to refine shared propagation paths improves estimates of entire radiation modes that supply no measurements. At 0~dB reference SNR, the proposed update reduced mean NMSE on two wholly unobserved modes by 11.37\% relative to ridge VarPro and by 3.49\% relative to adaptive ridge. Coefficient correction alone captured nearly nine tenths of the full-channel gain over VarPro. Correcting noise in the shared gains can improve extrapolation even when the observed residual rises.

\bibliographystyle{IEEEtran}\bibliography{references}
\end{document}


\raggedbottom\maketitle\thispagestyle{plain}
\begin{abstract}
This supplement develops the observation model and implementation of noise-aware path refinement for cross-pattern channel extrapolation. Fixed-geometry risk calculations motivate coefficient correction, and the update equations specify its noise scaling, physical constraints and symmetry properties. Paired experiments examine coefficient learning, deterministic shrinkage, unobserved modes and sensitivity to model departures. Complete numerical tables report the variation across channels and fitted networks. Additional implementation and timing results document the computational requirements and numerical precision of the estimator.
\end{abstract}

\section{Calibrated Channel and Acquisition Model}\label{supp:model}
\subsection{Propagation state and channel law}
\label{sec:system}
The propagation state is $s=\{(c_\ell,u_\ell,v_\ell)\}_{\ell=1}^L$, where $c_\ell\in\mathbb C$ is a path coefficient and $u_\ell,v_\ell$ are receive and transmit spatial frequencies. All modes share this state. The calibrated field gain $g_k$ determines how mode $k$ weights each arrival direction. The channel tensor $H=\R(s)\in\mathbb C^{K\times N_r\times N_t}$ has entries
\begin{equation}
 H_{krt}=\sum_{\ell=1}^{L}c_\ell g_k(u_\ell)
             e^{\mathrm i\pi(ru_\ell-tv_\ell)}.\label{eq:renderer}
\end{equation}
Both element indices start at zero and the steering vectors are unnormalized. The matched configuration fixes $N_r=64$, $N_t=8$, $K=4$ and $L=10$. The receive frequency lies in $[-\sin(\pi/3),\sin(\pi/3)]$. Transmit frequencies are represented on $[-1,1)$ with period two. The estimator retains ten path slots in every experiment except the explicitly identified true-order reference.

For each simulated channel, receive angles are independent uniform draws on $[-60^\circ,60^\circ]$ and transmit angles are independent uniform draws on $[-90^\circ,90^\circ]$. Their sines give $u_\ell$ and $v_\ell$. Independently, draw $z_\ell\sim\CN(0,1)$ and set $c=z/\|z\|_2$. Consequently, $\sum_\ell |c_\ell|^2=1$, while the channel energy varies with the angles, path interference and radiation gains. Coefficient normalization introduces dependence among the path magnitudes.

\subsection{Row-wise acquisition and unobserved modes}
In receive row $r$, one acquired mode supplies all transmit-column measurements. Its local index $o_r$ is mapped to target index $q(o_r)$ by matching the source descriptor to the target descriptor table. The measurements are
\begin{equation}
 y_{rt}=H_{q(o_r),rt}+n_{rt},\qquad
 n_{rt}\sim\CN(0,\sigma^2).
\end{equation}
The noise is independent across acquired entries. After vectorization, $y=SH+n$ with $S\in\{0,1\}^{m\times D}$ selecting $m=512$ complex entries from $D=2048$ target entries. Throughout the equations, $S\R(s)$ denotes the vector of acquired entries.

Balanced four-mode acquisition, denoted \texttt{all4} in the implementation, observes 16 rows in each mode. Balanced two-mode acquisition, denoted \texttt{pair03}, observes 32 rows in each of modes 0 and 3. Row assignments are randomly permuted for each acquisition. The latter mask leaves modes 1 and 2 entirely unobserved while retaining their calibrated response functions as inference inputs. Their propagation state is inferred from the observed modes and rendered through~\eqref{eq:renderer}. Both masks therefore estimate the same four-mode target from the same number of acquired entries.

The unobserved modes occur in the training distribution and their descriptors are available at test time. The extrapolation concerns missing observations for the current channel realization. The measurement count refers to complex channel entries. Mapping this count to pilot symbols requires a specified signaling and receiver implementation.

\subsection{Calibration representation}
A descriptor contains field-gain values and scaled slope samples at the 129 knots $u_j=-1+2j/128$. With $a_{\max}=10^{8/20}$, its entries decode as
\begin{equation}
 g_k(u_j)=a_{\max}d_{k,0,j},\qquad
 g'_k(u_j)=64a_{\max}d_{k,1,j}.
\end{equation}
Cubic Hermite interpolation reconstructs the gain between knots. The nominal descriptor uses centered finite differences at interior knots and one-sided differences at the endpoints. The nominal pattern family has offsets $(45,-45,15,-15)$ degrees and field amplitudes
\begin{equation}
 g_k(u)=10^{[8-\min\{12(\psi_k(u)/65)^2,30\}]/20},
\end{equation}
where $\psi_k(u)=\operatorname{wrap}((180/\pi)\arcsin u+\delta_k)$ is in degrees. VarPro evaluates this analytic function and its analytic derivative. Refinement evaluates the supplied Hermite interpolant and its derivative, including when a mismatch experiment perturbs the true propagation gains. The zero-stage control quantifies the effect of transferring the fitted state between these two representations.

\subsection{Reference and received SNR}
The reference power $P_0=0.9997065117346389$ is the mean energy per entry of the 5,000 training channels evaluated with unit radiation gain. This fixed value precedes pattern weighting and row selection. A reference SNR $\rho_{\rm ref}$ in dB sets $\sigma^2=P_0 10^{-\rho_{\rm ref}/10}$ for both masks. The realized received SNR is
\begin{equation}
 \rho_{\rm rx}=10\log_{10}\frac{\|SH\|_2^2}{m\sigma^2}.\label{eq:received_snr}
\end{equation}
The evaluator computes~\eqref{eq:received_snr} from the noiseless acquired entries. Inference receives $\sigma^2$, the row assignment and the descriptors. Unless explicitly identified as received SNR, every SNR value in this supplement uses the reference scale.

\section{Noise-Aware Path Refinement}\label{sec:refinement}
\subsection{Observation fit and complete-channel risk}\label{sec:risk}
At fixed path angles, collect the complete atoms in $B\in\mathbb C^{D\times L}$ and the observed atoms in $A=SB\in\mathbb C^{m\times L}$. Here $S$ is the row-selection operator defined in Section~\ref{sec:system}. For fixed $c$, assume $A$ has full column rank and $y=Ac+n$, with $n\sim\CN(0,\sigma^2I)$. For $\widehat c=(A^HA)^{-1}A^Hy$, a deterministic real $t$ gives
\begin{align}
R(t)&=\mathbb E\|Bt\widehat c-Bc\|^2=(1-t)^2E+t^2V,\\
E&=\|Bc\|^2,\qquad V=\sigma^2\operatorname{tr}[B(A^HA)^{-1}B^H].
\end{align}
The expectation over measurement noise removes the cross term. If $E,V>0$, $t_*=E/(E+V)$ gives $R(t_*)=EV/(E+V)<V$. For the same estimate, the expected error on the measured data is
\begin{equation}
\mathbb E\|y-At\widehat c\|^2=(m-L)\sigma^2+(1-t)^2(\|Ac\|^2+L\sigma^2),
\end{equation}
which is minimized at $t=1$. Orthogonality of the least-squares residual gives residual growth for every realized $y$ if $t\ne1$ and $AA^\dagger y\ne0$. The minimizing factor depends on unknown target energy. Its role is to explain how shrinkage can improve the reconstruction while increasing the observed residual.

This fixed-geometry calculation conditions on the atom matrices. Estimating the angles changes both atom matrices, and coherent atoms couple the path uncertainties through $A^HA$ and $B^HB$. Normalizing the Gaussian coefficient vector in our simulator also makes its components dependent.

\subsection{Risk after ridge regression}
The inference-time VarPro initializer uses ridge regression at each candidate geometry,
\begin{equation}
\widehat c_P=(A^HA+m\lambda_PI)^{-1}A^Hy,
\quad A^H(y-A\widehat c_P)=m\lambda_P\widehat c_P.
\end{equation}
For $\lambda_P>0$, a nonzero fitted coefficient and its backprojected residual share the same phase. Either nonzero complex response, with two real output weights, spans the coefficient plane. Combining the responses changes the range of bounded updates available at a given state. After an LM or learned step the state need no longer satisfy the ridge normal equations. Moving the initial state from the analytic gain function to the interpolated response also introduces a small numerical discrepancy. Section~\ref{sec:mechanism} measures these departures along the refinement trajectory.

For fixed $A$ and $Q=(A^HA+m\lambda_PI)^{-1}A^H$, the exact ridge risk decomposes as
\begin{equation}
\mathbb E\|BQy-Bc\|^2=\|B(QA-I)c\|^2+\sigma^2\operatorname{tr}(BQQ^HB^H).
\end{equation}
The ridge multiplier affects both terms through $Q$. The network can assign different updates to different paths, conditional on the observed data. The deterministic controls test how much of the gain can be recovered by shrinking weak coefficients while holding the angles fixed.

\subsection{Analytic and learned update}
\subsubsection{Initialization and residual linearization}
Both initializers fit the ten-path state using the acquired rows and the registered pattern responses. During training, Newtonized orthogonal matching pursuit (NOMP)~\cite{mamandipoor2016nomp} searches a $96\times48$ grid. Each detection is followed by one Newton step and one cyclic refinement round. Training uses the state after the last such round. The receive grid includes both endpoints and the transmit grid covers one half-open period. We chose this setting from three candidates on the first 16 development realizations. All training comparisons use the same initial state for each input. At evaluation, a greedy search on a $128\times64$ grid initializes ridge variable projection (VarPro)~\cite{golub1973varpro,golub2003varpro}, which fits the angles jointly and solves for the coefficients at each candidate geometry. The learned weights are trained on the NOMP states and transferred unchanged to VarPro states at inference. Training caches the fitted state using the data types, shapes and bytes of all five input arrays. Every timed inference call performs a fresh fit.

Let $x$ pack $[\Re c_\ell,\Im c_\ell,u_\ell,v_\ell]$ for each path. Let $[\Phi_\ell]_{krt}=g_k(u_\ell)e^{\mathrm i\pi(ru_\ell-tv_\ell)}$ denote a full-channel atom and let $S$ select the 512 observed entries. The measured Jacobian $J\in\mathbb C^{m\times4L}$ differentiates complex channel entries with respect to these real coordinates. Its block for path $\ell$ is
\begin{equation}
 J_\ell=S[\Phi_\ell,\,\mathrm i\Phi_\ell,\,c_\ell\partial_u \Phi_\ell,\,c_\ell\partial_v \Phi_\ell],
\end{equation}
where $\partial_u \Phi_\ell=[g'_k(u_\ell)+\mathrm i\pi r g_k(u_\ell)]e^{\mathrm i\pi(ru_\ell-tv_\ell)}$ and $\partial_v \Phi_\ell=-\mathrm i\pi t \Phi_\ell$. The derivative $g'_k$ is that of the supplied Hermite interpolant. Thus the real adjoint is $J^*_{\mathbb R}z=\Re(J^Hz)$, with no division by $g_k$ at a pattern zero.

For $e_j=\|J_{:j}\|_2^2$, the implementation sets
\begin{align}
 f_{\mathrm{LM}}&=10^{-12}\max\{\max_j e_j,\tau_{64}\},\\
 D_c&=\operatorname{diag}\big(\sqrt{\max(e_j,f_{\mathrm{LM}})}\big),
\end{align}
where $\tau_{64}$ is the smallest positive normal float64 value. The scaling is defined in the physical coordinates used by the channel model. On valid, nondegenerate inputs let $r=y-S\R(s)$ and $\bar J=JD_c^{-1}$. The fixed proposal is
\begin{align}
 G&=\Re(\bar J^H\bar J),& b&=\Re(\bar J^Hr),\\
 \mu&=0.01(1+\sigma^2/P_0),&
 \delta_{\mathrm{LM}}&=D_c^{-1}(G+\mu I)^{-1}b.\label{eq:lm}
\end{align}
We symmetrize the real Gram matrix and solve the 40-variable system by Cholesky factorization. The damping is dimensionless. Writing $z=(G+\mu I)^{-1}b$, we have $\langle J^*_{\mathbb R}r,\delta_{\mathrm{LM}}\rangle=b^Tz>0$ for $b\ne0$. Thus the analytic proposal gives a local descent direction for the unprojected observation loss. The algorithm applies the learned correction and projection before taking a finite step.

\subsubsection{Network and phase-equivariant coefficient bases}
Each stage concatenates 12 real features of one path with their means across all paths. A shared $24\to16\to16\to6$ MLP with SiLU activations maps these features to six real outputs. The features describe coefficient and angular sensitivity, gradient alignment, atom coherence, residual energy, stage and noise level. The coordinate intended for the query-Gram discrepancy is fixed at zero in all evaluated controllers. Section~\ref{sec:features} gives the full vector.

For each path, the six bounded outputs are $q_j=\tanh z_j$, $j=1,\ldots,6$. The final layer starts at zero, so the untrained network reproduces the fixed estimator on the same initial state. The 774 parameters are shared across paths and stages.

Let $a_\ell$ be the observed coefficient atom, $d_\ell=\|a_\ell\|^2$, $g_\ell=a_\ell^Hr$, and $\nu=\sigma^2+10^{-6}P_0$. Define
\begin{align}
 d_{\mathrm{floor}}&=10^{-12}\max(\max_\ell d_\ell,1),\\
 \widetilde d_\ell&=\max(d_\ell,d_{\mathrm{floor}}),&
 a_{c,\ell}&=\sqrt{\nu/\widetilde d_\ell},\\
 e_{c,\ell}&=\frac{c_\ell}{\sqrt{|c_\ell|^2+a_{c,\ell}^2}},&
 e_{g,\ell}&=\frac{g_\ell}{\sqrt{|g_\ell|^2+\nu\widetilde d_\ell}}.
\end{align}
The coefficient correction is
\begin{equation}
 \Delta c_\ell=\frac{a_{c,\ell}}{\sqrt2}
 [(q_1+\mathrm i q_2)e_{c,\ell}+(q_3+\mathrm i q_4)e_{g,\ell}].\label{eq:dc}
\end{equation}
Stable hypotenuse evaluations implement the two complex denominators. Since $a_{c,\ell}$ grows as the observed atom energy decreases, weakly observed paths receive a larger noise-scaled correction range. This scale is a feature of the update rule.

For observed angular Jacobian column energies $d_{u,\ell},d_{v,\ell}$ and training-grid spacings $h_u=\sqrt3/95$, $h_v=2/48$, set
\begin{align}
 a_{u,\ell}&=\sqrt{\frac{\nu}{d_{u,\ell}+\nu/h_u^2}},&
 a_{v,\ell}&=\sqrt{\frac{\nu}{d_{v,\ell}+\nu/h_v^2}},\\
 \Delta u_\ell&=a_{u,\ell}q_5,&
 \Delta v_\ell&=a_{v,\ell}q_6.
\end{align}
The spacings bound a single learned step. Continuous path coordinates can move across grid cells over successive stages. The total proposal is $\delta=\delta_{\mathrm{LM}}+\Delta x$.

\subsubsection{Acceptance rule and supervised objective}
For $\alpha\in\{1,1/2,\ldots,1/256\}$ in descending order, the algorithm projects $u$ to its domain, wraps $v$ modulo two, and evaluates the supplied calibrated channel model. Writing this projected candidate as $s^+$, it accepts the first scale satisfying
\begin{equation}
 \frac{\|\R(s^+)-\R(s)\|_F}{\sqrt D}
 \leq \frac{1}{2}\frac{\|y-S\R(s)\|_2}{\sqrt m}+10^{-12}.\label{eq:cap}
\end{equation}
If no candidate passes, the state stays fixed. A nonfinite proposal is recorded as a failure. During training, gradients follow the projected state and renderer for the selected scale. The discrete scale and the cached pursuit estimates are held fixed in the backward pass.

For a batch of $N_b$ channels, training minimizes full-channel NMSE after the third stage,
\begin{equation}
 \mathcal L=\frac1{N_b}\sum_{i=1}^{N_b}
       \frac{\|\R(s_{\theta,i}^{(3)})-H_i\|_F^2}{\max\{\|H_i\|_F^2,10^{-12}\}}.
\end{equation}
The target channel appears in the training objective and the evaluation metric. At inference, the final estimate is reconstructed as $\widehat H=\R(s^{(3)})$.

\subsubsection{Bounds and symmetries of the state update}
\begin{proposition}
For finite features and valid positive scales, the added corrections satisfy $|\Delta c_\ell|\leq2a_{c,\ell}$, $|\Delta u_\ell|\leq h_u$, and $|\Delta v_\ell|\leq h_v$.
\end{proposition}
\begin{proof}
Both complex bases have modulus at most one, and each pair $q_i+\mathrm i q_j$ has modulus at most $\sqrt2$. The triangle inequality proves the coefficient bound. The angular denominators are at least $\nu/h^2$, and $|q_j|\leq1$.
\end{proof}
The bound applies to the learned term. The combined LM and learned proposal is further restricted by the channel-increment cap in~\eqref{eq:cap}.

\begin{proposition}
For a fixed descriptor and observation operator, assume the features and scales are finite. Under a common global phase rotation of the observations and coefficient state, the added coefficient correction rotates by the same phase, while the angular correction is invariant. A common permutation of the estimated path triples permutes the corrections.
\end{proposition}
\begin{proof}
The coefficient and gradient bases acquire the same phase. The real features, column energies and damping are invariant. A path permutation reorders the features and the four-column Jacobian blocks while leaving the path mean fixed. Equal scaling of the real and imaginary coefficient columns commutes with their joint rotation. Projection, wrapping and the rendered-increment test preserve both transformations in exact arithmetic.
\end{proof}
The result assumes the corresponding initial states. A discrete pursuit can change support near a tie, and floating-point rounding can alter an acceptance decision at the threshold.

When $b=0$, the analytic direction in~\eqref{eq:lm} vanishes. Equation~\eqref{eq:dc} can still produce a correction through $c_\ell$. At zero residual, however,~\eqref{eq:cap} permits at most $10^{-12}$ RMS change in the rendered channel. If both $c_\ell$ and $g_\ell$ vanish, the learned coefficient term for that path is zero. Coupling through LM can still change the coefficient within the ten-slot model.

The acceptance rule controls movement in channel space. A scalar example illustrates its effect on the observed residual. For $y=1$, a current estimate of zero and a step of $-0.1$ satisfy the cap while increasing the squared residual from 1 to 1.21. The coherence normalization in Section~\ref{sec:features} requires nondegenerate observed atoms. With an identically zero atom, multiplying binary64 energy floors can underflow before the square root. The implementation rejects resulting nonfinite features. The nominal positive-gain patterns satisfy this nondegeneracy requirement.

\subsection{Feature construction and numerical scaling}
\label{sec:features}
Each path contributes twelve real features. The first seven encode fitted strength, backprojected residual and angular sensitivity. The remaining features supply coherence, residual and acquisition context. Let $a_\ell=S\Phi_\ell$, $d_\ell=\|a_\ell\|_2^2$, $p_\ell=|c_\ell|^2$, $g_\ell=a_\ell^Hr$, $b_{u,\ell}=\Re[(S c_\ell\partial_u \Phi_\ell)^Hr]$, and analogously $b_{v,\ell}$. Let $d_{u,\ell},d_{v,\ell}$ denote the squared norms of the corresponding angular Jacobian columns, and $\nu=\sigma^2+10^{-6}P_0$. With $\epsilon_{\rm tiny}$ the smallest positive normal binary64 value, the features are
\begin{align}
 f_{\ell,0}&=\log(1+p_\ell d_\ell/\nu),\\
 f_{\ell,1}&=\log(1+|g_\ell|^2/\max(d_\ell\nu,\epsilon_{\rm tiny})),\\
 f_{\ell,2}&=\Re(c_\ell^*g_\ell)/(|c_\ell||g_\ell|+\nu),\\
 f_{\ell,3}&=\operatorname{asinh}(b_{u,\ell}/\sqrt{\max(d_{u,\ell}\nu,\epsilon_{\rm tiny})}),\\
 f_{\ell,4}&=\operatorname{asinh}(b_{v,\ell}/\sqrt{\max(d_{v,\ell}\nu,\epsilon_{\rm tiny})}),\\
 f_{\ell,5}&=\log\!\left(1+\frac{d_{u,\ell}}{\pi^2N_r^2(p_\ell d_\ell+\nu)}\right),\\
 f_{\ell,6}&=\log\!\left(1+\frac{d_{v,\ell}}{\pi^2N_t^2(p_\ell d_\ell+\nu)}\right),\\
 f_{\ell,7}&=\max_{j\ne\ell}|(R_o)_{\ell j}|,\qquad f_{\ell,8}=0,\\
 f_{\ell,9}&=\log(1+\|r\|_2^2/(m\nu)),\\
 f_{\ell,10}&=j/\max(T-1,1),\\
 f_{\ell,11}&=\log(1+\sigma^2/P_0).
\end{align}
Here $j=0,1,2$ and $T=3$. The observed normalized atom Gram matrix is $(R_o)_{ij}=a_i^Ha_j/\sqrt{e_ie_j}$ with $e_i=\max(d_i,\epsilon_{\rm tiny})$. Coordinate 8 is zero in every model. The controller input is $[f_\ell,\bar f]\in\mathbb R^{24}$, where $\bar f=L^{-1}\sum_\ell f_\ell$. All coordinates can be computed from the measurements and the current state.

The three numerical floors serve distinct purposes. The coefficient basis floors atom energy at $10^{-12}\max(1,\max_jd_j)$. The joint LM solve uses the column scaling specified above, while feature normalization uses the binary64 tiny value. These choices are fixed across all matched learned methods.

Including biases, the three layers allocate $(24\cdot16+16)+(16\cdot16+16)+(16\cdot6+6)=774$ real parameters. The zero local and pooled query-Gram features leave 32 first-layer weights inactive. Masking the two angular outputs in the coefficient-only model leaves 34 output-layer parameters inactive. The noise/stage MLP retains only the local and pooled stage and supplied-variance features, leaving 320 first-layer weights attached to masked coordinates. All architectures retain the same allocated network dimensions. In single-response retraining, two coefficient outputs are masked and both angular outputs remain available.

\section{Training and Control Estimators}\label{sec:training}
\subsection{Inference procedure}
The estimator decodes the mask, fits ten paths with VarPro, evaluates their coefficients and frequencies with the Hermite-interpolated gains, and executes three refinement stages. Its output is the four-mode channel reconstructed directly from the final path state. Section~\ref{sec:inputs} specifies the input arrays and arithmetic precision.

\subsection{Training data and optimization}

The five predefined seeds, numbered in the order used in Table~\ref{tab:individual}, are 20261102, 20261103, 2026110901, 2026110902, and 2026110903. Each arm uses the same 5000 training realizations and 30000 fixed acquisition views. The condition cycle is $(\texttt{all4},0)$, $(\texttt{pair03},0)$, $(\texttt{all4},10)$, $(\texttt{pair03},10)$, $(\texttt{all4},20)$, $(\texttt{pair03},20)$. For seed $z$, each condition uses PCG64 with SeedSequence $[z,5000,\text{source index},\text{SNR},0]$ to permute all training indices. Update $t$ takes condition $t\bmod6$ and its next eight indices. Thus every view occurs once in 3750 updates.

Adam uses learning rate $0.001$, betas $(0.9,0.999)$, epsilon $10^{-8}$ and zero weight decay. We clip the global gradient norm at one, use float64/complex128 arithmetic in training and disable AMP, TF32 and AMSGrad. All arms start from identical weights for a given seed. We evaluate the final checkpoint for each fit.

We cache training-time NOMP estimates by hashes of the five input arrays. The targets enter only the NMSE loss. We used 500 separate development channels to choose the architecture and primary baselines.

\subsection{Ablations and physical controls}
The analytic controls set all learned outputs to zero and run three or twelve stages. A zero-stage control measures transfer from the analytic gains to the Hermite model. The coefficient-only network masks $q_5,q_6$, so only LM moves the angles. The noise/stage network masks all features except $f_{10},f_{11}$ before concatenating local and pooled inputs. This removes the measurement residual feature as well as the path-specific features, while retaining the network dimensions and physical response bases.

The coefficient-ridge and joint-quadratic controls use coupled Gram matrices and a coefficient penalty with scale $\lambda=\kappa L\sigma^2/P_0$. Development selected $\kappa=3$ for both. They retain the same three-stage transition and channel-increment cap. Section~\ref{sec:quadratic} gives their full update equations and the associated selection grids. The adaptive-ridge control in Section~\ref{sec:shrinkage} instead holds all angles fixed while updating its diagonal penalty from the current coefficient energies.

\onecolumn
\section{Matched-Channel Confirmation Study}
\label{sec:closure}
\subsection{Pairing and uncertainty intervals}
The primary estimators, checkpoints, comparisons and sample size were fixed before the draw of 1,152 independent ten-path channels. Their full-channel hashes were distinct from those of 11,848 previously registered channels. Each test channel has one view for each mask at each of three SNRs. The underlying channel and full-array noise draw are common to the two masks at a given SNR. Each estimator receives the same row assignments and noisy values as its competitors.

Each view was evaluated with six deterministic estimators and five fitted versions of each of three learned estimators (full, coefficient-only and noise/stage). The resulting 6,912 acquisition views produced 145,152 finite predictions, with 21 estimators or fitted models per view. Repeatability checks are reported in Section~\ref{sec:numerics}.

For channel $i$ and fitted model $z$, define $e_{iz}=\|\widehat H_{iz}-H_i\|_F^2/\|H_i\|_F^2$. Each learned method contributes the fit mean $e_i=\sum_z e_{iz}/5$ for that channel. The reported risk is the arithmetic mean $\bar e$ across channels, computed in linear units. For a reference $b$, the reported reduction is $100(1-\bar e/\bar e_b)\%$. This ratio of means can differ from the mean or median of the per-channel reductions.

At 0~dB, the full estimator had three prespecified comparisons per mask, against VarPro, coefficient-only refinement and the noise/stage network. A common set of 100000 paired-channel bootstrap draws gave two-sided 99.1667\% percentile intervals for the ratios of mean errors, with a Bonferroni adjustment over the six comparisons. Other intervals are descriptive 95\% intervals. These intervals measure variation across simulated channels conditional on the five fits. The individual-fit table reports variation across training seeds. Averaging the errors of five fitted networks quantifies their typical performance and does not require averaging their predicted channels at deployment.

\subsection{Magnitude and distribution of the error}
Table~\ref{tab:newprimary} gives the prespecified ratios. Tables~\ref{tab:means}--\ref{tab:tails} give absolute errors, individual fits and distribution tails for the same channels. Figure~\ref{fig:absolute_nmse} locates these errors on the scale of the observation-matched direct and linear estimates.
\begin{resulttable}\centering\small
\caption{Six prespecified comparisons at 0~dB reference SNR. Ratios divide the mean NMSE of full refinement by that of the named reference. Values below one favor full refinement. The 99.1667\% intervals use a Bonferroni adjustment for the six comparisons.}
\label{tab:newprimary}
\begin{tabular}{@{}llrr@{}}\toprule\rowcolor{TableHeader}
Mask & Reference & Ratio & Interval\\\midrule
Four & VarPro & 0.8852 & [0.8738, 0.8967]\\
Four & Coefficient-only & 0.9865 & [0.9839, 0.9892]\\
Four & Noise/stage & 0.9069 & [0.8972, 0.9167]\\
Two & VarPro & 0.8827 & [0.8716, 0.8941]\\
Two & Coefficient-only & 0.9863 & [0.9837, 0.9890]\\
Two & Noise/stage & 0.9055 & [0.8958, 0.9154]\\
\bottomrule\end{tabular}\end{resulttable}
\begin{resulttable}\centering\small
\caption{Mean full-channel NMSE ($\times10^{-3}$) across 1,152 channels. Columns give reference SNR in dB. Learned rows average the errors of five fitted models within each channel. Bold entries mark the lowest mean in each column. The shaded row identifies full refinement.}
\label{tab:means}
\begin{tabular}{@{}lrrrrrr@{}}\toprule\rowcolor{TableHeader}
 & \multicolumn{3}{c}{Four modes sampled} & \multicolumn{3}{c}{Two modes sampled}\\
\cmidrule(lr){2-4}\cmidrule(lr){5-7}
Method & 0 dB & 10 dB & 20 dB & 0 dB & 10 dB & 20 dB\\\midrule
VarPro & 17.59847 & 1.42533 & 0.13613 & 19.89240 & 1.54037 & 0.14832\\
Zero-step Hermite & 17.59847 & 1.42533 & 0.13613 & 19.89240 & 1.54037 & 0.14832\\
Analytic (3 stages) & 17.87674 & 1.42682 & 0.13610 & 20.26094 & 1.54319 & 0.14828\\
Analytic (12 stages) & 17.88711 & 1.42720 & \textbf{0.13609} & 20.26408 & 1.54353 & \textbf{0.14826}\\
Coefficient ridge & 17.67202 & 1.42949 & 0.13623 & 19.97655 & 1.54482 & 0.14838\\
Joint quadratic & 17.61768 & 1.42557 & 0.13615 & 19.91109 & 1.54154 & 0.14834\\
\rowcolor{TableFocus} Full & \textbf{15.57758} & \textbf{1.39031} & 0.13646 & \textbf{17.55866} & \textbf{1.50521} & 0.14854\\
Coefficient-only & 15.79116 & 1.39657 & 0.13675 & 17.80196 & 1.51165 & 0.14885\\
Noise/stage MLP & 17.17671 & 1.42180 & 0.13618 & 19.39104 & 1.53742 & 0.14833\\
\bottomrule\end{tabular}\end{resulttable}
\begin{resulttable}\centering\small
\caption{Individual-fit mean NMSE ($\times10^{-3}$), retaining every predefined training seed. The seed numbering is $1=20261102$, $2=20261103$, $3=2026110901$, $4=2026110902$, $5=2026110903$.}
\label{tab:individual}
\begin{tabular}{@{}lrrrrrrr@{}}\toprule\rowcolor{TableHeader}
 & & \multicolumn{3}{c}{Four modes sampled} & \multicolumn{3}{c}{Two modes sampled}\\
\cmidrule(lr){3-5}\cmidrule(lr){6-8}
Method & Seed & 0 dB & 10 dB & 20 dB & 0 dB & 10 dB & 20 dB\\\midrule
\addlinespace[2pt]
Full & 1 & 15.54450 & 1.38870 & 0.13613 & 17.48711 & 1.50353 & 0.14825\\
Full & 2 & 15.51468 & 1.38741 & 0.13566 & 17.52284 & 1.50195 & 0.14764\\
Full & 3 & 15.53591 & 1.38851 & 0.13627 & 17.54249 & 1.50367 & 0.14834\\
Full & 4 & 15.54448 & 1.39167 & 0.13719 & 17.54213 & 1.50720 & 0.14918\\
Full & 5 & 15.74835 & 1.39528 & 0.13704 & 17.69875 & 1.50971 & 0.14929\\
\addlinespace[2pt]
Coefficient-only & 1 & 15.68898 & 1.39470 & 0.13644 & 17.67827 & 1.50965 & 0.14855\\
Coefficient-only & 2 & 15.66854 & 1.39040 & 0.13580 & 17.68704 & 1.50520 & 0.14780\\
Coefficient-only & 3 & 15.78387 & 1.39235 & 0.13573 & 17.81702 & 1.50781 & 0.14774\\
Coefficient-only & 4 & 15.70007 & 1.39575 & 0.13760 & 17.71093 & 1.51127 & 0.14971\\
Coefficient-only & 5 & 16.11436 & 1.40965 & 0.13816 & 18.11653 & 1.52431 & 0.15047\\
\addlinespace[2pt]
Noise/stage MLP & 1 & 17.17233 & 1.42151 & 0.13618 & 19.37968 & 1.53707 & 0.14833\\
Noise/stage MLP & 2 & 17.17340 & 1.42158 & 0.13619 & 19.38239 & 1.53713 & 0.14835\\
Noise/stage MLP & 3 & 17.17677 & 1.42178 & 0.13608 & 19.39989 & 1.53742 & 0.14825\\
Noise/stage MLP & 4 & 17.18300 & 1.42221 & 0.13635 & 19.39112 & 1.53789 & 0.14848\\
Noise/stage MLP & 5 & 17.17804 & 1.42192 & 0.13608 & 19.40215 & 1.53760 & 0.14825\\
\bottomrule\end{tabular}\end{resulttable}

\begin{resulttable}\small
\caption{Distribution of full-channel NMSE across 1,152 confirmation channels ($\times10^{-3}$). Percentiles and maxima were computed after averaging learned errors over fits within each channel.}
\label{tab:tails}
\begin{tabular}{@{}lr rrrr rrrr@{}}\toprule
 & & \multicolumn{4}{c}{Four modes sampled} & \multicolumn{4}{c}{Two modes sampled}\\
\cmidrule(lr){3-6}\cmidrule(lr){7-10}
Method & Ref. dB & Median & $p_{95}$ & $p_{99}$ & Max. & Median & $p_{95}$ & $p_{99}$ & Max.\\\midrule
VarPro & 0 & 16.4 & 29.51 & 35.98 & 42.78 & 18.22 & 35.9 & 47 & 57.68\\
 & 10 & 1.364 & 2.244 & 2.89 & 3.809 & 1.464 & 2.518 & 3.101 & 4.382\\
 & 20 & 0.1328 & 0.1955 & 0.2357 & 0.3721 & 0.1414 & 0.2265 & 0.2738 & 0.5315\\
\addlinespace[2pt]
Zero-step Hermite & 0 & 16.4 & 29.51 & 35.98 & 42.78 & 18.22 & 35.9 & 47 & 57.68\\
 & 10 & 1.364 & 2.244 & 2.89 & 3.809 & 1.464 & 2.518 & 3.101 & 4.382\\
 & 20 & 0.1328 & 0.1955 & 0.2357 & 0.3721 & 0.1414 & 0.2265 & 0.2738 & 0.5315\\
\addlinespace[2pt]
Analytic (3 stages) & 0 & 16.6 & 30.56 & 36.65 & 44.5 & 18.54 & 36.36 & 47.55 & 58.55\\
 & 10 & 1.363 & 2.224 & 2.901 & 3.84 & 1.465 & 2.505 & 3.096 & 4.352\\
 & 20 & 0.1325 & 0.1944 & 0.2361 & 0.3719 & 0.1416 & 0.2273 & 0.2742 & 0.5328\\
\addlinespace[2pt]
Analytic (12 stages) & 0 & 16.6 & 30.58 & 36.68 & 44.43 & 18.51 & 36.33 & 47.57 & 58.55\\
 & 10 & 1.364 & 2.233 & 2.901 & 3.84 & 1.464 & 2.505 & 3.096 & 4.378\\
 & 20 & 0.1325 & 0.1938 & 0.2361 & 0.3715 & 0.1416 & 0.2272 & 0.2742 & 0.5331\\
\addlinespace[2pt]
Coefficient ridge & 0 & 16.44 & 29.79 & 36.17 & 42.45 & 18.26 & 36.25 & 47.04 & 58.87\\
 & 10 & 1.367 & 2.263 & 2.888 & 3.805 & 1.471 & 2.528 & 3.091 & 4.731\\
 & 20 & 0.1329 & 0.1956 & 0.2356 & 0.3755 & 0.1413 & 0.2276 & 0.273 & 0.5331\\
\addlinespace[2pt]
Joint quadratic & 0 & 16.39 & 29.49 & 36.02 & 42.57 & 18.26 & 35.66 & 46.93 & 57.65\\
 & 10 & 1.366 & 2.244 & 2.889 & 3.804 & 1.465 & 2.525 & 3.085 & 4.392\\
 & 20 & 0.1327 & 0.1956 & 0.2356 & 0.3721 & 0.1413 & 0.2264 & 0.2735 & 0.5314\\
\addlinespace[2pt]
Full & 0 & 14.88 & 24.51 & 30.1 & 37.61 & 16.3 & 29.52 & 39.68 & 55.28\\
 & 10 & 1.351 & 2.086 & 2.489 & 3.309 & 1.45 & 2.375 & 2.782 & 3.455\\
 & 20 & 0.1336 & 0.1965 & 0.2328 & 0.3615 & 0.1424 & 0.2246 & 0.2598 & 0.4704\\
\addlinespace[2pt]
Coefficient-only & 0 & 15.13 & 25.16 & 30.34 & 35.99 & 16.49 & 30.08 & 40.25 & 55.49\\
 & 10 & 1.356 & 2.084 & 2.502 & 3.376 & 1.453 & 2.391 & 2.84 & 3.497\\
 & 20 & 0.134 & 0.1967 & 0.2334 & 0.3606 & 0.1426 & 0.2259 & 0.2622 & 0.4821\\
\addlinespace[2pt]
Noise/stage MLP & 0 & 16.2 & 28.17 & 34.98 & 39.3 & 17.83 & 34.29 & 44.97 & 55.67\\
 & 10 & 1.362 & 2.243 & 2.85 & 3.77 & 1.465 & 2.502 & 3.056 & 4.312\\
 & 20 & 0.1328 & 0.1956 & 0.2344 & 0.3731 & 0.1415 & 0.2259 & 0.2744 & 0.531\\
\bottomrule\end{tabular}\end{resulttable}

At 20~dB the full estimator had higher fit-averaged error than VarPro on 721 of 1152 channels with the four-mode mask and on 701 with the two-mode mask. Both intervals for the ratio of mean errors included one.

\begin{resultfigure}\centering
\includegraphics[width=\textwidth]{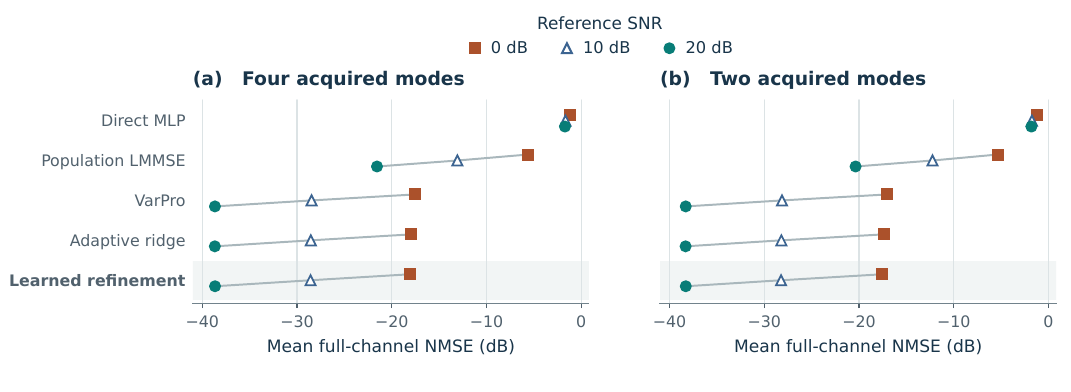}
\caption{Mean full-channel NMSE on 1,152 channels for four-mode acquisition (a) and two-mode acquisition (b), including observation-matched direct and linear estimators. Values in dB are $10\log_{10}$ of the arithmetic mean in linear units.}\label{fig:absolute_nmse}
\end{resultfigure}

\begin{resultfigure}\centering
\includegraphics[width=\textwidth]{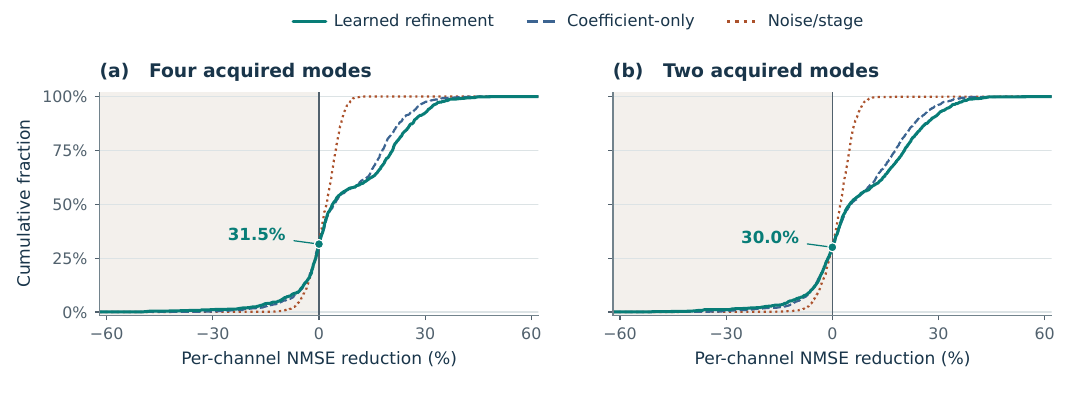}
\caption{Cumulative distribution of per-channel NMSE reductions from VarPro at 0~dB. Negative values denote channels with increased error. The labeled points at zero give their fraction for full refinement. Learned errors are averaged over fits within each channel.}\label{fig:scene_distribution}
\end{resultfigure}

Figure~\ref{fig:scene_distribution} shows how often a positive aggregate gain translates to improvement on an individual channel. At 0~dB the median per-channel reductions were 3.84\% for the four-mode mask and 4.95\% for the two-mode mask. The learned correction increased error on 363 and 346 of the 1152 channels, respectively, but lowered the 95th percentile of full-channel error by 16.95\% and 17.77\%. We sorted the channels for each mask by the initial VarPro coefficient energy $\sum_\ell|\widehat c_\ell|^2$ (288 per quartile). From the lowest quartile to the highest, the reductions in mean NMSE were 12.46, 11.45, 9.59 and 12.23\% for the four-mode mask and 12.93, 11.94, 9.85 and 11.81\% for the two-mode mask. The fraction with higher error rose from 20.8\% and 17.0\% in the bottom quartiles to 37.5\% and 37.8\% in the top quartiles. Mean improvement did not vary monotonically with coefficient energy in this post-hoc analysis.

Figure~\ref{fig:paired_contrasts} compares mean errors at each noise level. The intervals use a common set of channels within each comparison.
\begin{resultfigure}
\includegraphics[width=\textwidth]{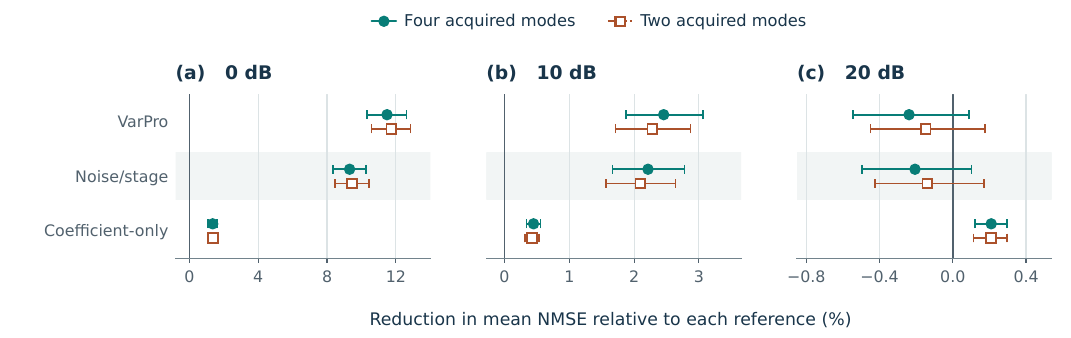}
\caption{Reduction in mean NMSE of full refinement against each reference on 1,152 channels. Each reference SNR has its own horizontal scale. Paired-channel intervals are 99.1667\% at 0~dB and descriptive 95\% elsewhere, conditional on the five fits.}\label{fig:paired_contrasts}
\end{resultfigure}

\subsection{Errors in measured and unmeasured entries}
\begin{resultfigure}\centering
\includegraphics[width=\textwidth]{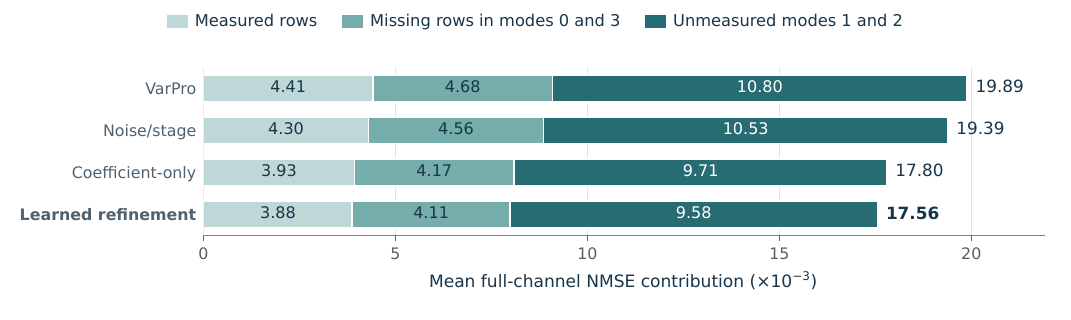}
\caption{Contributions of three disjoint channel regions to full-channel NMSE under two-mode acquisition at 0~dB. All regions in a channel share its full-energy denominator. Labels give mean contributions in thousandths. Each row sums to the full-channel mean.}\label{fig:regional_error}
\end{resultfigure}
Figure~\ref{fig:regional_error} separates the initial error by both row coverage and mode coverage. For the two-mode mask at 0~dB, VarPro's NMSE separates into 0.00441227 on measured entries, 0.00467657 on unmeasured entries in modes 0 and 3, and 0.01080356 in modes 1 and 2. Each term uses the full-channel energy denominator. The learned correction lowered the contribution of the entirely unobserved modes to 0.009577. Relative to the VarPro contribution from those same modes, the reduction was 11.35\%.

\subsection{Error normalized within the unobserved modes}
The contribution decomposition above weights each region by the energy of the complete channel. To measure extrapolation accuracy within the entirely unobserved modes, let $\mathcal U=\{1,2\}$ and define
\begin{equation}
 e_{i,\mathcal U}=\frac{\sum_{k\in\mathcal U}\|\widehat H_{i,k}-H_{i,k}\|_F^2}
 {\sum_{k\in\mathcal U}\|H_{i,k}\|_F^2}.\label{eq:unobserved_nmse}
\end{equation}
Both numerator and denominator are restricted to $\mathcal U$ before averaging across channels. The distinction changes the weighting of channels according to their energy in the unobserved modes.

We reran VarPro, the frozen adaptive-ridge rule and all five learned models on the original 1,152 two-mode views at 0~dB. Each method used the same observations, calibration descriptors and noise variance. The trained weights and reconstruction settings were fixed. Table~\ref{tab:unobserved} reports the resulting errors. The learned estimator reduced mean unobserved-mode NMSE by 11.37\% relative to VarPro and by 3.49\% relative to adaptive ridge. Its fit-averaged error was lower on 797 and 787 of the 1,152 channels, respectively.

\begin{resulttable}\small
\caption{NMSE restricted to the entirely unobserved modes at 0~dB reference SNR. The denominator in each channel is the energy of modes 1 and 2. Reductions compare full refinement with the reference in each row. Intervals are descriptive 95\% paired-channel intervals from 10,000 draws, conditional on the five fitted networks.}
\label{tab:unobserved}
\begin{tabular}{@{}lrrr@{}}\toprule
Estimator & Mean NMSE ($\times10^{-3}$) & Full-refinement gain (\%) & Channels improved\\\midrule
VarPro & 21.4560 & 11.37 [10.46, 12.29] & 797 / 1152\\
Adaptive ridge & 19.7028 & 3.49 [3.01, 3.96] & 787 / 1152\\
\rowcolor{TableFocus} Full refinement & \textbf{19.0160} & \multicolumn{2}{c}{---}\\
\bottomrule\end{tabular}\end{resulttable}
This post-hoc reanalysis uses the confirmation population and complements the additive regional contributions in Fig.~\ref{fig:regional_error}.

\section{Mechanisms of the Learned Correction}\label{sec:mechanism}

\subsection{Trajectories from the VarPro estimate}
After the test results were available, we traced the first 128 test channels in their original order, using both masks, three SNRs and all five learned fits. The recorded trajectories reproduced the original final outputs to absolute tolerance $10^{-12}$ and relative tolerance $10^{-11}$.

Table~\ref{tab:stages} reports the learned coefficient alignment $\Re(c_\ell^*\Delta c_\ell)$ alongside diagnostics of the accepted total change, which includes the LM step, scaling and projection. We measured the residual and full-channel error before and after each stage. For a given mask and SNR the 128 independent channels supplied 1920 stage executions and 19200 path updates across the five fits.

\begin{resulttable}\centering\small
\caption{Trajectories from VarPro on the first 128 confirmation channels and all five fits. Opposing counts learned coefficient increments with negative alignment. The fourth column counts accepted total stages that increase the residual while reducing full-channel error. Cosines compare the phases of the two complex bases, pooled over paths. States $s^{(0)},s^{(1)},s^{(2)}$ enter stages 1, 2, and 3, respectively.}\label{tab:stages}
\begin{tabular}{@{}llrrrrr@{}}\toprule\rowcolor{TableHeader}
Mask & SNR & Opposing & Residual up, & \multicolumn{3}{c}{$\cos(e_c,e_g)$ at incoming state}\\ & (dB) & paths & risk down & $s^{(0)}$ & $s^{(1)}$ & $s^{(2)}$ \\\midrule
Four & 0 & 99.93\% & 43.02\% & 1.0000 & 0.9166 & 0.9762\\
Four & 10 & 99.94\% & 25.05\% & 1.0000 & 0.9657 & 0.9762\\
Four & 20 & 99.81\% & 23.85\% & 1.0000 & 0.9577 & 0.8565\\
Two & 0 & 99.76\% & 43.33\% & 1.0000 & 0.9124 & 0.9761\\
Two & 10 & 99.96\% & 28.33\% & 1.0000 & 0.9646 & 0.9760\\
Two & 20 & 99.82\% & 25.73\% & 1.0000 & 0.9598 & 0.8599\\
\bottomrule\end{tabular}\end{resulttable}

In Fig.~\ref{fig:stage_diagnostics}, the first stage lowers the average channel error even though many accepted updates increase the measurement residual. At that stage the median cosine of the phase difference between coefficient and gradient was approximately one. We also computed $\|A^Hr-m\lambda_Pc\|/\|m\lambda_Pc\|$ using the Hermite-interpolated gains. Its small initial value reflects the change in gain representation. It grows as subsequent updates move the coefficients away from the starting ridge solution. The learned term also tends to oppose the coefficient at high SNR, where mean NMSE is close to that of VarPro. The sign of the update alone therefore gives little information about its effect on the estimate.

\begin{resultfigure}\centering
\includegraphics[width=\textwidth]{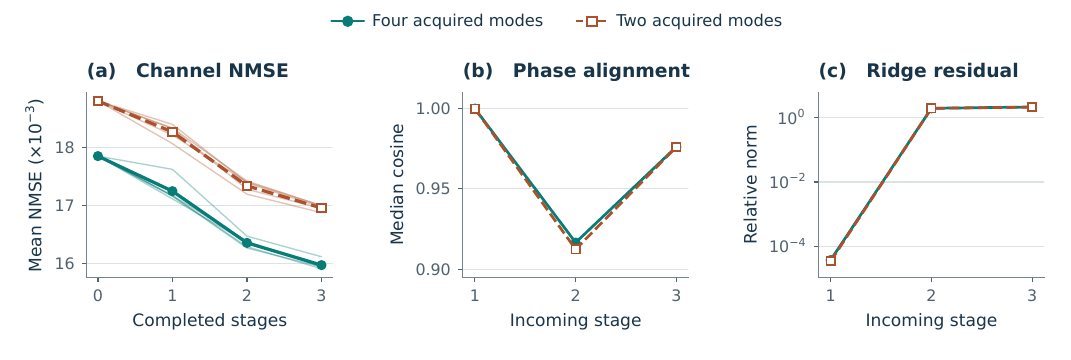}
\caption{VarPro-initialized refinement on the first 128 confirmation channels at 0~dB. (a) Mean full-channel NMSE after each stage. Thin lines denote single fits and thick lines their mean. (b,c) Median phase cosine and normalized ridge residual on entry to each stage. The ridge residual uses a log scale.}\label{fig:stage_diagnostics}
\end{resultfigure}

\subsection{Coefficient responses and update capacity}
To separate the two phase references, we retrained the controller once with the coefficient response $e_c$ alone and once with the backprojected residual response $e_g$ alone. Both retained the learned angle outputs and the analytic LM step. Table~\ref{tab:bases} reports their means and Table~\ref{tab:single_fits} gives every fitted model. For each of five seeds the two models used the same initialization, NOMP training estimates, 30000 views, minibatch order and 3750 Adam updates. We evaluated the ten resulting fits after VarPro on the 1152 previously tested channels (69120 predictions).

At 0~dB, $e_c$ alone reduced mean error from VarPro by 8.21\% and 8.18\% for the four- and two-mode masks, respectively. The corresponding reductions for $e_g$ alone were 8.26\% and 8.28\%, compared with 11.48\% and 11.73\% for the mixed response. At 10~dB the reductions were 1.86\% and 1.77\% with $e_c$, and 1.78\% and 1.65\% with $e_g$. At 20~dB, $e_c$ alone gave slightly lower mean error than the mixed response.

For each mask and SNR we averaged errors across the five fits within a channel. The intervals in Table~\ref{tab:single_ratios} use 10000 paired-channel bootstrap resamples (seed 2026111007). They are descriptive and conditional on these fits.

Masking one complex output pair changes the maximum coefficient increment from $2a_c$ to $a_c$. One nonzero response with two real weights already spans the local coefficient plane. A single-response controller therefore changes the available phase information and tightens the update bound simultaneously. The repeated-response controls below restore the common uniform bound.

\begin{resulttable}\centering\small
\caption{Single-response controls. Mean NMSE ($\times10^{-3}$) averages five fitted models within each channel. Both controls retain the learned angle updates.}
\label{tab:bases}
\begin{tabular}{@{}llrrr@{}}\toprule\rowcolor{TableHeader}
Acquisition & Basis & 0 dB & 10 dB & 20 dB\\\midrule
Four & $e_c$ only & 16.1530 & 1.3988 & 0.1361\\
Four & $e_g$ only & 16.1453 & 1.4000 & 0.1367\\
Two & $e_c$ only & 18.2650 & 1.5130 & 0.1481\\
Two & $e_g$ only & 18.2452 & 1.5149 & 0.1488\\
\bottomrule\end{tabular}\end{resulttable}

\begin{resulttable}\centering\small
\caption{Ratios of single-response to full mean NMSE with descriptive 95\% paired-channel intervals. Values above one favor full refinement. All five final fits are used.}\label{tab:single_ratios}
\begin{tabular}{@{}llccc@{}}\toprule\rowcolor{TableHeader}
Acquisition & Basis & 0 dB & 10 dB & 20 dB\\\midrule
Four & $e_c$ only & 1.0369 [1.0331, 1.0407] & 1.0061 [1.0041, 1.0081] & 0.9973 [0.9962, 0.9985]\\
Four & $e_g$ only & 1.0364 [1.0335, 1.0393] & 1.0070 [1.0054, 1.0085] & 1.0015 [1.0006, 1.0024]\\
Two & $e_c$ only & 1.0402 [1.0363, 1.0440] & 1.0052 [1.0033, 1.0071] & 0.9972 [0.9961, 0.9983]\\
Two & $e_g$ only & 1.0391 [1.0362, 1.0420] & 1.0064 [1.0049, 1.0080] & 1.0016 [1.0007, 1.0025]\\
\bottomrule\end{tabular}\end{resulttable}
\begin{resulttable}\centering\small
\caption{Mean NMSE ($\times10^{-3}$) for each retrained single-response controller on the 1152 test channels.}\label{tab:single_fits}
\begin{tabular}{@{}llrrrrrr@{}}\toprule\rowcolor{TableHeader}
& & \multicolumn{3}{c}{Four} & \multicolumn{3}{c}{Two}\\
Basis & Seed & 0 dB & 10 dB & 20 dB & 0 dB & 10 dB & 20 dB\\\midrule
$e_c$ only & 20261102 & 15.9693 & 1.3969 & 0.1362 & 18.0643 & 1.5112 & 0.1482\\
$e_c$ only & 20261103 & 16.4888 & 1.4043 & 0.1359 & 18.6281 & 1.5194 & 0.1479\\
$e_c$ only & 2026110901 & 15.9500 & 1.3941 & 0.1357 & 18.0437 & 1.5079 & 0.1477\\
$e_c$ only & 2026110902 & 15.9697 & 1.3980 & 0.1363 & 18.0622 & 1.5121 & 0.1483\\
$e_c$ only & 2026110903 & 16.3874 & 1.4006 & 0.1363 & 18.5269 & 1.5146 & 0.1484\\
$e_g$ only & 20261102 & 16.0629 & 1.3992 & 0.1368 & 18.1507 & 1.5157 & 0.1489\\
$e_g$ only & 20261103 & 16.0393 & 1.3968 & 0.1371 & 18.1413 & 1.5102 & 0.1493\\
$e_g$ only & 2026110901 & 16.1438 & 1.3984 & 0.1362 & 18.2427 & 1.5152 & 0.1483\\
$e_g$ only & 2026110902 & 16.2828 & 1.4027 & 0.1360 & 18.3950 & 1.5162 & 0.1480\\
$e_g$ only & 2026110903 & 16.1978 & 1.4029 & 0.1371 & 18.2965 & 1.5173 & 0.1493\\
\bottomrule\end{tabular}\end{resulttable}
\subsubsection{Controlling for output count and update magnitude}
We also trained controllers that repeat $e_c$ or $e_g$ in both complex terms of the coefficient update. Each run retained all six network outputs, including the two angle outputs, the $1/\sqrt{2}$ normalization and the $2a_c$ uniform coefficient bound. We used the same five seeds and NOMP training schedules as above. Table~\ref{tab:repeated} reports the complete repeated-response results. Because the two coefficient terms now receive the same response, their zero-initialized output heads have exchange symmetry. Repeating a response preserves the uniform bound while changing the increments available at a given state.

\begin{resulttable}\small
\caption{Repeated-response controls. Gain is the percentage reduction in mean NMSE relative to VarPro. A ratio above one indicates higher risk than the mixed-response full controller.}\label{tab:repeated}
\begin{tabular}{@{}lrllrrr@{}}\toprule
Mask & Ref. dB & Responses & Mean NMSE & Gain (\%) & Ratio to full & 95\% ratio interval \\
\midrule
Four & 0 & $e_c,e_c$ & 0.015878 & 9.77 & 1.0193 & [1.0164, 1.0221] \\
Four & 0 & $e_g,e_g$ & 0.015758 & 10.46 & 1.0116 & [1.0102, 1.0130] \\
Four & 10 & $e_c,e_c$ & 0.0013943 & 2.18 & 1.0029 & [1.0014, 1.0044] \\
Four & 10 & $e_g,e_g$ & 0.001392 & 2.34 & 1.0012 & [1.0004, 1.0021] \\
Four & 20 & $e_c,e_c$ & 0.00013617 & -0.03 & 0.9979 & [0.9968, 0.9990] \\
Four & 20 & $e_g,e_g$ & 0.00013668 & -0.40 & 1.0016 & [1.0009, 1.0024] \\
Two & 0 & $e_c,e_c$ & 0.017927 & 9.88 & 1.0210 & [1.0182, 1.0237] \\
Two & 0 & $e_g,e_g$ & 0.017825 & 10.39 & 1.0152 & [1.0135, 1.0169] \\
Two & 10 & $e_c,e_c$ & 0.0015091 & 2.03 & 1.0026 & [1.0011, 1.0041] \\
Two & 10 & $e_g,e_g$ & 0.0015072 & 2.15 & 1.0013 & [1.0004, 1.0022] \\
Two & 20 & $e_c,e_c$ & 0.0001482 & 0.08 & 0.9977 & [0.9966, 0.9988] \\
Two & 20 & $e_g,e_g$ & 0.00014875 & -0.29 & 1.0014 & [1.0007, 1.0022] \\
\bottomrule\end{tabular}\end{resulttable}

In Fig.~\ref{fig:matchedseeds}, the mixed response gives lower 0-dB mean error than either repeated response for every matched seed under both masks. Repeating $e_g$ lowered VarPro error by 10.46\% and 10.39\%. Relative to the same VarPro denominator, the mixed response gained a further 1.03 and 1.34 percentage points. Relative to the repeated-$e_g$ estimator itself, this is a 1.15\% and 1.49\% reduction in mean error. At 20~dB, repeating $e_c$ gave the lowest mean error of the three.
\begin{resultfigure}\centering
\includegraphics[width=\textwidth]{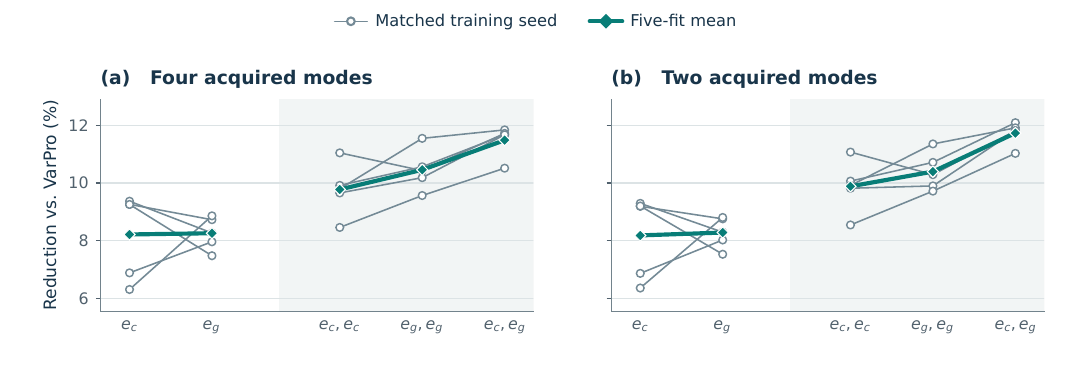}
\caption{Reduction in mean NMSE relative to VarPro at 0~dB on 1,152 channels. Thin lines connect fits with the same training seed within each architecture group. The heavy line gives the five-fit mean. Repeated-response and mixed-response controllers have the same parameter count.}\label{fig:matchedseeds}
\end{resultfigure}

\subsection{Comparison with explicit coefficient shrinkage}\label{sec:shrinkage}
We compared 24 estimators using 32 additional development channels under both masks at 0~dB. Besides native VarPro, we tested projection onto the unit coefficient ball, least squares constrained to that ball, seven pathwise shrinkage rules and fourteen coupled adaptive ridge rules. The unit radius follows from the generator. For atom $a_\ell$, let $t_\ell=\sigma^2/\max(\|a_\ell\|^2,10^{-15})$. The pathwise rule rescales the fitted coefficient to $c_\ell|c_\ell|^2/(|c_\ell|^2+\kappa t_\ell)$. Adaptive ridge retains the complete Gram matrix $A^HA$ and starts from the VarPro coefficient vector $c^{(0)}$. With the angles fixed, it iterates
\begin{equation}
 \big[A^HA+\operatorname{diag}_{\ell}\!\{\kappa\sigma^2/\max(|c_\ell^{(j)}|^2+t_\ell,10^{-30})\}\big]c^{(j+1)}=A^Hy.
\end{equation}
Both searches used $\kappa\in\{0.1,0.3,1,3,10,30,100\}$, with one or three solves for adaptive ridge. Three solves with $\kappa=3$ minimized mean development error across the two masks. We fixed this choice before evaluating the 64 additional channels.
\begin{resulttable}\small
\caption{Post-hoc deterministic controls on 1152 original realizations. Values give reduction in mean complete-channel NMSE (\%) from VarPro. The final column uses adaptive ridge as its denominator and gives a descriptive 95\% paired-channel interval.}\label{tab:shrinkage}
\begin{tabular}{@{}lrrrrrrl@{}}\toprule
Mask & Ref. dB & Projection & Ball LS & Gate & Adaptive & Full & Full/adaptive [95\% CI] \\
\midrule
Four & 0 & -0.05 & 0.42 & 4.14 & 8.14 & 11.48 & 3.64 [3.23, 4.04] \\
Four & 10 & -4.82 & -4.08 & 0.96 & 1.73 & 2.46 & 0.74 [0.52, 0.96] \\
Four & 20 & -28.98 & -26.39 & 0.12 & 0.24 & -0.24 & -0.48 [-0.64, -0.33] \\
Two & 0 & 0.05 & 0.54 & 3.89 & 8.42 & 11.73 & 3.62 [3.19, 4.04] \\
Two & 10 & -4.38 & -3.85 & 0.77 & 1.54 & 2.28 & 0.76 [0.52, 0.99] \\
Two & 20 & -36.70 & -34.23 & 0.17 & 0.34 & -0.15 & -0.49 [-0.64, -0.34] \\
\bottomrule\end{tabular}\end{resulttable}

\section{Generalization across Noise Levels}\label{sec:noise}
The reference power $P_0=0.9997065117346389$ is the mean of $\|A_r\operatorname{diag}(c)A_t^H\|_F^2/(64\cdot8)$ over the 5000 training channels evaluated with unit radiation gain. The same $P_0$ sets the noise variance for every four-mode simulation. Actual received SNR uses the noiseless power on the measured rows. At 0~dB reference SNR, the central 90\% of received SNRs across the 1152 test channels was [4.03,5.35]~dB for the four-mode mask and [3.64,5.46]~dB for the two-mode mask.
\subsection{Noise sweep on 64 additional channels}
The 64 test channels are distinct from both the development channels and the original 1152 test channels. All estimators share each set of observations. The paired bootstrap uses 10000 draws to give pointwise descriptive intervals conditional on the five fits. Four reference SNRs, $-10,-5,5,15$~dB, were absent from training.
\begin{resulttable}\small
\caption{Independent noise sweep on 64 channels. Received SNR is the median and central 90\% range of per-channel received SNR (dB). Errors are linear mean NMSE ($\times10^{-3}$). The final column gives the percentage reduction of full relative to adaptive ridge.}\label{tab:noisesweep}
\begin{tabular}{@{}lrlrrrl@{}}\toprule
Mask & Ref. dB & Received SNR & VarPro & Adaptive & Full & Reduction [95\% CI] \\
\midrule
Four & -10 & -5.23 [-5.89, -4.79] & 273.5 & 212.2 & 213.7 & -0.68 [-5.15, 3.30] \\
Four & -5 & -0.23 [-1.07, 0.24] & 74.65 & 62.58 & 60.69 & 3.02 [0.23, 5.62] \\
Four & 0 & 4.78 [4.00, 5.27] & 18.45 & 16.59 & 15.98 & 3.69 [2.01, 5.31] \\
Four & 5 & 9.80 [9.03, 10.14] & 4.757 & 4.56 & 4.493 & 1.49 [0.15, 2.78] \\
Four & 10 & 14.80 [13.95, 15.29] & 1.451 & 1.428 & 1.424 & 0.23 [-0.85, 1.27] \\
Four & 15 & 19.77 [19.07, 20.31] & 0.4662 & 0.462 & 0.4649 & -0.63 [-1.92, 0.57] \\
Four & 20 & 24.72 [24.10, 25.30] & 0.1301 & 0.1302 & 0.1313 & -0.91 [-1.35, -0.47] \\
Two & -10 & -5.21 [-6.51, -4.62] & 290.7 & 229 & 235.4 & -2.82 [-7.04, 1.41] \\
Two & -5 & -0.23 [-1.44, 0.35] & 83.88 & 69.07 & 65.24 & 5.54 [2.53, 8.18] \\
Two & 0 & 4.78 [3.47, 5.37] & 21.25 & 19.07 & 18.31 & 4.02 [2.39, 5.71] \\
Two & 5 & 9.80 [8.41, 10.43] & 5.223 & 5.02 & 4.921 & 1.97 [0.73, 3.22] \\
Two & 10 & 14.71 [13.40, 15.39] & 1.59 & 1.563 & 1.542 & 1.36 [0.69, 2.10] \\
Two & 15 & 19.74 [18.57, 20.41] & 0.4696 & 0.4656 & 0.465 & 0.13 [-0.55, 0.95] \\
Two & 20 & 24.81 [23.28, 25.37] & 0.1514 & 0.151 & 0.151 & 0.03 [-0.45, 0.59] \\
\bottomrule\end{tabular}\end{resulttable}
At $-5$~dB reference SNR, each of the five fits improved on adaptive ridge. Their individual reductions ranged from 2.18\% to 4.04\% for the four-mode mask and from 4.67\% to 7.01\% for the two-mode mask. At $-10$~dB only one fit had lower mean error than adaptive ridge for each mask. We also timed the two masks at the three training SNRs on the first two channels, giving twelve common single-thread CPU inputs. We rotated method order and timed three calls per input after three warmups. Each call included a fresh path fit and excluded model loading. Taking medians first over repetitions and then over inputs gave 539.57~ms for VarPro, 541.91~ms with adaptive ridge and 610.12~ms with the learned correction.

\section{Direct and Linear Channel Estimators}\label{sec:baselines}
\subsection{A linear estimator with population covariance}
The zero-mean matched channel has covariance $C=C_r\otimes C_t$ when modes and receive rows precede transmit columns in the stacked vector. Unit total coefficient energy, independent angles, and zero pairwise coefficient cross-moments yield
\begin{align}
[C_r]_{(k,r),(j,v)}&=\mathbb E_{\theta}\!\left[g_k(\sin\theta)g_j(\sin\theta)^*
 e^{\mathrm j\pi(r-v)\sin\theta}\right],\\
[C_t]_{tw}&=J_0\!\left(\pi(t-w)\right).
\end{align}
Here $\theta$ is uniform on $[-60^\circ,60^\circ]$. The Bessel term follows from the uniform transmit angle on $[-90^\circ,90^\circ]$. The estimator computes $CS^H(SCS^H+\sigma^2I)^{-1}y$ using the known mask, noise variance and matched angular law. We evaluate the receive covariance with 144 Gauss--Legendre nodes per smooth pattern interval. Doubling the order changed any entry by at most $3.6\times10^{-14}$. On a two-mode example, the factorized implementation agreed with a dense 512-dimensional solve to within $1.1\times10^{-14}$. Using the exact population covariance gives the minimum expected squared error over linear estimators for the matched distribution. This optimality concerns unnormalized squared error under the matched distribution. The sparse estimators additionally fit the path parameters of each realization.

\subsection{Complex-valued direct-estimation controls}
The direct estimator is an adaptation of the complex network of Liang and Li~\cite{liang2024prnet} to our observation operator. Their PR-Net uses grouped transmit-side acquisition and an LMMSE input estimator. Their stated simulation uses eight transmit-side modes, training at 25 dB and 500 epochs. We retain three 512-unit complex hidden layers, independent ReLU activations on real and imaginary parts, a complex linear readout, and Adam (batch 32, learning rate 0.001). Here reconfiguration is on the receive side, the query has four registered modes, and the row mask varies. The input combines 2,048 zero-filled complex channel entries, a 256-entry row-to-mode mask and the log noise variance. The descriptors map source owners into the known query ordering. Each fit has 5,513,216 real parameters. The network predicts a residual added to the zero-filled observation. It trains on full-channel NMSE for four passes through the same 30000 views, giving 3752 updates. By comparison, the 774-parameter controller uses one pass and 3750 updates at batch eight. Thus the two models see the same views and take nearly equal numbers of optimizer steps. Batch size, total view presentations, compute and structural priors differ.
The direct control uses the final fourth-epoch checkpoint for each of the five predefined seeds. A separate variant fixed to the first seed replaces the measured entries in its input by their population-LMMSE filtered values and adds the learned residual to the complete population-LMMSE estimate. That front end knows the additional matched angular law. Both controls are scored on the 1152 original realizations and the 64 new noise-sweep realizations. No score from these panels selects a checkpoint.
\begin{resulttable}\small
\caption{Observation-matched alternatives. Entries are mean complete-channel NMSE in linear units. Direct MLP averages five errors per scene. MLP + LMMSE uses the first fixed seed. Full refinement also averages five errors per scene. Bold entries identify the lowest mean for each observation setting.}\label{tab:direct}
\begin{tabular}{@{}llrrrrrr@{}}\toprule
Population & Mask & Ref. dB & Direct MLP & MLP + LMMSE & Pop. LMMSE & VarPro & Full \\
\midrule
1152 & Four & 0 & 0.7656 & 0.2751 & 0.2751 & 0.0176 & \textbf{0.01558} \\
1152 & Four & 10 & 0.683 & 0.04939 & 0.04936 & 0.001425 & \textbf{0.00139} \\
1152 & Four & 20 & 0.6742 & 0.007017 & 0.006992 & \textbf{0.0001361} & 0.0001365 \\
\addlinespace[3pt]
1152 & Two & 0 & 0.752 & 0.2907 & 0.2907 & 0.01989 & \textbf{0.01756} \\
1152 & Two & 10 & 0.6698 & 0.05982 & 0.05979 & 0.00154 & \textbf{0.001505} \\
1152 & Two & 20 & 0.6612 & 0.009234 & 0.00921 & \textbf{0.0001483} & 0.0001485 \\
\addlinespace[3pt]
64 & Four & -10 & 1.606 & 0.7216 & 0.7215 & 0.2735 & \textbf{0.2137} \\
64 & Four & -5 & 0.9631 & 0.5013 & 0.5013 & 0.07465 & \textbf{0.06069} \\
64 & Four & 0 & 0.7624 & 0.2786 & 0.2786 & 0.01845 & \textbf{0.01598} \\
64 & Four & 5 & 0.698 & 0.1244 & 0.1244 & 0.004757 & \textbf{0.004493} \\
64 & Four & 10 & 0.6762 & 0.05004 & 0.04998 & 0.001451 & \textbf{0.001424} \\
64 & Four & 15 & 0.6702 & 0.01871 & 0.01868 & 0.0004662 & \textbf{0.0004649} \\
64 & Four & 20 & 0.6678 & 0.007041 & 0.00701 & \textbf{0.0001301} & 0.0001313 \\
\addlinespace[3pt]
64 & Two & -10 & 1.597 & 0.736 & 0.736 & 0.2907 & \textbf{0.2354} \\
64 & Two & -5 & 0.9498 & 0.5156 & 0.5155 & 0.08388 & \textbf{0.06524} \\
64 & Two & 0 & 0.7499 & 0.3012 & 0.3012 & 0.02125 & \textbf{0.01831} \\
64 & Two & 5 & 0.6878 & 0.1458 & 0.1458 & 0.005223 & \textbf{0.004921} \\
64 & Two & 10 & 0.6659 & 0.06125 & 0.06122 & 0.00159 & \textbf{0.001542} \\
64 & Two & 15 & 0.6595 & 0.02378 & 0.02376 & 0.0004696 & \textbf{0.000465} \\
64 & Two & 20 & 0.6569 & 0.009298 & 0.009272 & 0.0001514 & \textbf{0.000151} \\
\bottomrule\end{tabular}\end{resulttable}
On 24 two-thread CPU inputs the median of per-input median times is 1.82~ms for the direct complex network and 3.15~ms with its LMMSE front end, including tensor construction and the final output array. The covariance is computed offline. The accompanying data give each fitted model's error on each scene. The training code regenerates the direct models.

\section{Controlled Departures from the Nominal Model}
\label{sec:stress}
\subsection{Experimental design}
We tested path count, gain calibration and channel power on 128 additional channels. We fixed nine conditions, two masks, three reference SNRs and the analysis before generating this sample. The 6912 views yielded 41472 predictions from the unchanged ten-slot VarPro estimator and the five learned fits.

For each channel we drew fourteen angle pairs and circular complex Gaussian coefficients from the nominal distributions. We retained the first $L$ paths, $L\in\{6,8,10,12,14\}$, and normalized the retained coefficients to unit squared norm for each $L$. This pairs the different path counts while preserving the marginal channel law at each count. Both estimators retain ten slots in this comparison and receive the same five inputs.

For the two calibration interventions, true $L=10$ and
\begin{equation}
H^{\mathrm{true}}_{krt}=10^{s_a a_k/20}\exp(\mathrm j\pi s_p p_k/180)\,\R(s)_{krt}.
\end{equation}
We independently drew two four-vectors of standard normals, subtracted their respective mode means and divided by $\sqrt{3/4}$ to obtain $a$ and $p$. Each component then has unit marginal variance. We set the gain and phase standard deviations $(s_a,s_p)$ to $(0.5\text{ dB},5^\circ)$ or $(1\text{ dB},10^\circ)$. Centering removes a common complex gain that could be absorbed into the path coefficients. The perturbation varies by mode and is constant in angle within a channel. We used the perturbed gains to generate measurements and targets and supplied the nominal gains to the estimators.

The power tests multiply all ten coefficients by $\sqrt{0.5}$ or $\sqrt{2}$ at fixed noise variance and $P_0$. This changes the received SNR while preserving the labels of 0, 10 and 20~dB on the reference scale.

We paired row assignments and full-array noise draws across conditions at a given reference SNR. We averaged per-channel errors across the five fits and used 10000 paired-channel bootstrap resamples (seed 2026111011) for descriptive 95\% percentile intervals. The resampling units are the 128 independent channels.

\subsection{Effect of model departures}
Figure~\ref{fig:mismatch} summarizes the 54 combinations of departure, mask and SNR. At 0~dB the ten-slot learned estimator reduced mean error from ten-slot VarPro by 55.08\% and 51.56\% for six true paths and by 40.56\% and 36.79\% for eight. These reductions persisted across the three tested SNRs. For every tested SNR and both masks, the fit-averaged error fell on all 128 channels when the true order was six or eight. Such behavior is consistent with suppressing noise fitted by excess paths.

When $L=12$, the 0-dB gains were 0.65\% and $-0.09\%$. When $L=14$, they were approximately 0\% and $-0.25\%$. At 20~dB the full estimator still had mean errors of about 0.019 and 0.021 for $L=12$ and 0.051 and 0.055 for $L=14$, compared with about $1.4\times10^{-4}$ and $1.5\times10^{-4}$ for $L=10$. The learned update cannot add slots when the model omits physical paths.

For calibration errors with standard deviations of 0.5 dB and $5^\circ$, full refinement reduced 0-dB mean NMSE by 6.21\% [4.20,8.17]\% and 4.89\% [3.06,6.63]\%. At 1 dB/$10^\circ$, they are 3.51\% [2.52,4.53]\% and 1.60\% [0.76,2.46]\%. The absolute NMSE of full refinement rose from 0.01565 to 0.05275 under four-mode acquisition and from 0.01762 to 0.07325 under two-mode acquisition. High-SNR gains are small or unresolved. Thus a relative gain can coexist with a substantial increase in absolute reconstruction error.

At half channel power the 0-dB reductions were 16.68\% and 14.69\%. At twice the power they were 6.60\% and 6.28\%. At 20~dB, both power settings and the nominal setting showed small negative gains. At nominal power and 0~dB the reductions on these 128 channels were 9.67\% and 10.09\%.

\begin{resultfigure}\centering
\includegraphics[width=\textwidth]{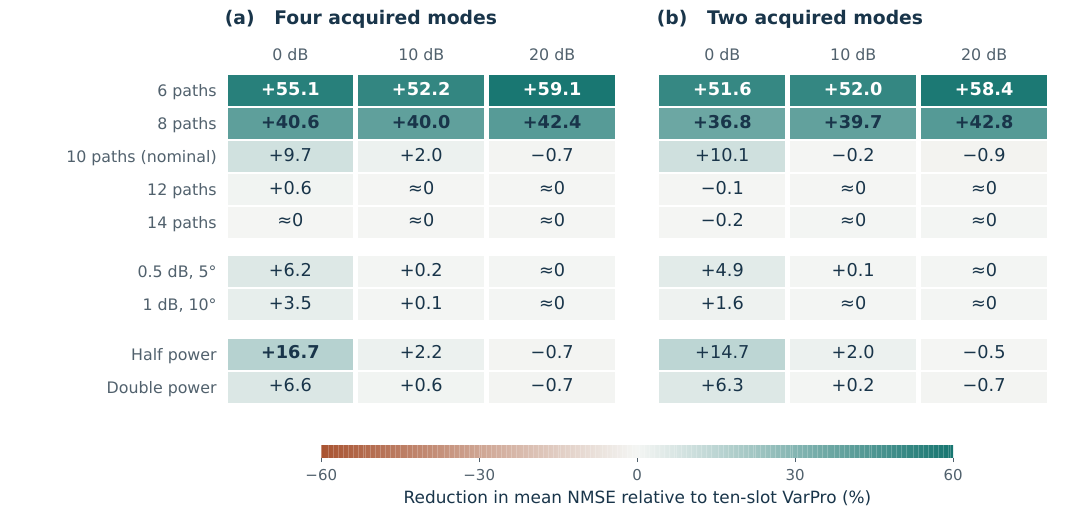}
\caption{Reduction in mean NMSE (\%) relative to ten-slot VarPro across all 54 conditions. Each cell uses the same 128 paired channels. Positive values favor full refinement. Calibration rows give the gain and phase standard deviations. The color scale is symmetric about zero. Values marked $\approx0$ have absolute reduction below 0.05\%. Tables~\ref{tab:stress_four} and \ref{tab:stress_two} report the underlying values and intervals.}\label{fig:mismatch}
\end{resultfigure}

\subsection{Complete numerical results}
\begin{resulttable}\centering\small
\caption{Controlled departures under four-mode acquisition. NMSE is in linear units. Gain and paired-channel intervals are percentages. The last column counts scenes with higher five-fit mean error than VarPro, out of 128.}\label{tab:stress_four}
\begin{tabular}{@{}lrccrc@{}}\toprule\rowcolor{TableHeader}
Setting & SNR & VarPro NMSE & Full NMSE & Gain [95\% interval] & Worse\\\midrule
$L=6$ & 0 & 0.0274009 & 0.0123081 & 55.081 [53.810, 56.361] & 0\\
  & 10 & 0.00281938 & 0.00134654 & 52.240 [50.961, 53.535] & 0\\
  & 20 & 0.000283117 & 0.000115739 & 59.120 [57.847, 60.401] & 0\\
\addlinespace[1.5pt]
$L=8$ & 0 & 0.022019 & 0.0130878 & 40.561 [39.300, 41.787] & 0\\
  & 10 & 0.00221016 & 0.00132561 & 40.022 [38.956, 41.062] & 0\\
  & 20 & 0.000217805 & 0.000125375 & 42.437 [41.155, 43.713] & 0\\
\addlinespace[1.5pt]
$L=10$ (nominal) & 0 & 0.0173205 & 0.0156451 & 9.673 [6.654, 12.505] & 51\\
  & 10 & 0.00143022 & 0.00140223 & 1.957 [0.658, 3.415] & 61\\
  & 20 & 0.000138264 & 0.000139192 & -0.671 [-1.027, -0.309] & 80\\
\addlinespace[1.5pt]
$L=12$ & 0 & 0.033837 & 0.0336176 & 0.648 [0.083, 1.366] & 60\\
  & 10 & 0.020621 & 0.0206236 & -0.013 [-0.038, 0.014] & 67\\
  & 20 & 0.0191252 & 0.0191277 & -0.013 [-0.020, -0.006] & 82\\
\addlinespace[1.5pt]
$L=14$ & 0 & 0.0663223 & 0.0663241 & -0.003 [-0.144, 0.180] & 70\\
  & 10 & 0.0531439 & 0.0531467 & -0.005 [-0.017, 0.007] & 71\\
  & 20 & 0.0514451 & 0.051446 & -0.002 [-0.007, 0.003] & 80\\
\addlinespace[1.5pt]
0.5 dB / $5^\circ$ & 0 & 0.0266499 & 0.0249946 & 6.211 [4.201, 8.167] & 49\\
  & 10 & 0.0107016 & 0.010675 & 0.249 [0.077, 0.459] & 61\\
  & 20 & 0.0094475 & 0.00944677 & 0.008 [-0.020, 0.034] & 62\\
\addlinespace[1.5pt]
1 dB / $10^\circ$ & 0 & 0.0546752 & 0.0527549 & 3.512 [2.519, 4.528] & 44\\
  & 10 & 0.0381779 & 0.0381373 & 0.106 [0.064, 0.151] & 50\\
  & 20 & 0.0368059 & 0.036793 & 0.035 [0.021, 0.049] & 48\\
\addlinespace[1.5pt]
Power $0.5\times$ & 0 & 0.0397532 & 0.0331215 & 16.682 [14.329, 18.823] & 26\\
  & 10 & 0.00284109 & 0.00277759 & 2.235 [0.931, 3.698] & 56\\
  & 20 & 0.000276585 & 0.000278581 & -0.721 [-1.189, -0.283] & 82\\
\addlinespace[1.5pt]
Power $2\times$ & 0 & 0.00776791 & 0.00725536 & 6.598 [3.877, 9.301] & 42\\
  & 10 & 0.00069563 & 0.000691717 & 0.563 [-0.499, 1.703] & 65\\
  & 20 & $6.91498\!\times\!10^{-5}$ & $6.96516\!\times\!10^{-5}$ & -0.726 [-1.061, -0.395] & 83\\
\bottomrule\end{tabular}\end{resulttable}
\begin{resulttable}\centering\small
\caption{Controlled departures under two-mode acquisition. NMSE is in linear units. Gain and paired-channel intervals are percentages. The last column counts scenes with higher five-fit mean error than VarPro, out of 128.}\label{tab:stress_two}
\begin{tabular}{@{}lrccrc@{}}\toprule\rowcolor{TableHeader}
Setting & SNR & VarPro NMSE & Full NMSE & Gain [95\% interval] & Worse\\\midrule
$L=6$ & 0 & 0.0296836 & 0.0143782 & 51.562 [49.691, 53.450] & 0\\
  & 10 & 0.00294056 & 0.00141125 & 52.008 [50.150, 53.754] & 0\\
  & 20 & 0.000299568 & 0.000124697 & 58.374 [57.081, 59.695] & 0\\
\addlinespace[1.5pt]
$L=8$ & 0 & 0.0246351 & 0.0155725 & 36.787 [35.142, 38.370] & 0\\
  & 10 & 0.00226941 & 0.00136904 & 39.674 [38.286, 41.084] & 0\\
  & 20 & 0.000231879 & 0.000132748 & 42.751 [41.165, 44.344] & 0\\
\addlinespace[1.5pt]
$L=10$ (nominal) & 0 & 0.019599 & 0.0176209 & 10.093 [7.350, 12.664] & 41\\
  & 10 & 0.00146459 & 0.00146824 & -0.249 [-1.917, 1.220] & 64\\
  & 20 & 0.000146799 & 0.000148177 & -0.939 [-1.919, -0.242] & 77\\
\addlinespace[1.5pt]
$L=12$ & 0 & 0.036811 & 0.0368438 & -0.089 [-0.866, 0.658] & 77\\
  & 10 & 0.0223293 & 0.0223319 & -0.012 [-0.034, 0.011] & 68\\
  & 20 & 0.0209528 & 0.0209546 & -0.009 [-0.015, -0.002] & 76\\
\addlinespace[1.5pt]
$L=14$ & 0 & 0.070242 & 0.0704158 & -0.247 [-0.406, -0.103] & 81\\
  & 10 & 0.0570311 & 0.0570309 & 0.000 [-0.011, 0.012] & 62\\
  & 20 & 0.0554495 & 0.0554485 & 0.002 [-0.004, 0.008] & 64\\
\addlinespace[1.5pt]
0.5 dB / $5^\circ$ & 0 & 0.0332908 & 0.0316623 & 4.892 [3.063, 6.625] & 51\\
  & 10 & 0.015745 & 0.0157249 & 0.128 [-0.015, 0.290] & 61\\
  & 20 & 0.0144531 & 0.0144487 & 0.031 [-0.024, 0.087] & 58\\
\addlinespace[1.5pt]
1 dB / $10^\circ$ & 0 & 0.0744451 & 0.0732519 & 1.603 [0.757, 2.462] & 53\\
  & 10 & 0.0574187 & 0.0573924 & 0.046 [-0.010, 0.102] & 54\\
  & 20 & 0.0561681 & 0.0561428 & 0.045 [0.007, 0.083] & 54\\
\addlinespace[1.5pt]
Power $0.5\times$ & 0 & 0.0447852 & 0.0382064 & 14.690 [12.231, 16.997] & 23\\
  & 10 & 0.00299934 & 0.00293921 & 2.005 [0.566, 3.657] & 53\\
  & 20 & 0.000295736 & 0.000297078 & -0.454 [-1.111, 0.163] & 77\\
\addlinespace[1.5pt]
Power $2\times$ & 0 & 0.0086658 & 0.00812193 & 6.276 [3.935, 8.523] & 44\\
  & 10 & 0.000730862 & 0.000729318 & 0.211 [-0.691, 1.158] & 58\\
  & 20 & $7.34381\!\times\!10^{-5}$ & $7.39461\!\times\!10^{-5}$ & -0.692 [-1.085, -0.285] & 84\\
\bottomrule\end{tabular}\end{resulttable}

\subsection{A true-order-informed physical reference}
For a post-hoc comparison we supplied the true path count to VarPro at $L=6$ and $L=8$. We retained its original grid, ridge multiplier and optimization budget. This comparison gives VarPro information unavailable to the learned ten-slot estimator. It covers both masks at all three SNRs (1536 predictions).

VarPro with the true count gave lower mean error in all twelve conditions. At 0~dB, the learned ten-slot estimates had 38.18\% and 38.95\% higher mean error for six true paths and 4.63\% and 7.78\% higher mean error for eight. The results in Table~\ref{tab:trueorder} show that coefficient refinement recovers only part of the error caused by excess fitted slots.

\begin{resulttable}\centering\small
\caption{Post-hoc true-order-informed reference. All NMSE values are linear. Ratios above one favor the reference with privileged order information. Descriptive 95\% intervals use 10000 paired-scene resamples, seed 2026111012.}\label{tab:trueorder}
\begin{tabular}{@{}llrccc@{}}\toprule\rowcolor{TableHeader}
True order & Mask & SNR & True-order VarPro & Ten-slot full & Full/reference [95\%]\\\midrule
6 & Four & 0 & 0.0089073 & 0.0123081 & 1.382 [1.316, 1.452]\\
6 & Four & 10 & 0.000845232 & 0.00134654 & 1.593 [1.533, 1.656]\\
6 & Four & 20 & $8.38169\!\times\!10^{-5}$ & 0.000115739 & 1.381 [1.348, 1.415]\\
6 & Two & 0 & 0.0103475 & 0.0143782 & 1.390 [1.311, 1.477]\\
6 & Two & 10 & 0.000871202 & 0.00141125 & 1.620 [1.544, 1.705]\\
6 & Two & 20 & $8.73328\!\times\!10^{-5}$ & 0.000124697 & 1.428 [1.384, 1.477]\\
8 & Four & 0 & 0.0125089 & 0.0130878 & 1.046 [1.007, 1.086]\\
8 & Four & 10 & 0.00114044 & 0.00132561 & 1.162 [1.132, 1.193]\\
8 & Four & 20 & 0.000110633 & 0.000125375 & 1.133 [1.119, 1.148]\\
8 & Two & 0 & 0.0144488 & 0.0155725 & 1.078 [1.036, 1.122]\\
8 & Two & 10 & 0.00116564 & 0.00136904 & 1.174 [1.146, 1.206]\\
8 & Two & 20 & 0.000116273 & 0.000132748 & 1.142 [1.127, 1.158]\\
\bottomrule\end{tabular}\end{resulttable}

\section{Transfer beyond the Nominal Configuration}\label{sec:additional}
\subsection{Transfer to other channel and pattern families}
An earlier transfer experiment used 32 realizations each of clustered delay line models CDL-B and CDL-C. Errors were averaged over three reference SNRs (0, 10 and 20~dB) and two fitted networks within a channel, then over channels. Under four- and two-mode acquisition, full refinement gave mean NMSE values of 0.742 and 0.744 on CDL-B, and 0.482 and 0.469 on CDL-C. At the reported precision, native VarPro gave 0.743 and 0.746 on CDL-B, and 0.481 and 0.467 on CDL-C. Both estimators therefore incurred substantial absolute error on these channel models. In a separate development test with a wider family of pattern descriptors (Shared V2), the strong initializer rejected the descriptors in all 192 views and returned no estimates.

\subsection{Sensitivity to the supplied noise variance}
A second two-fit development test held the observations fixed on 32 matched channels and supplied half, the correct value or twice the correct noise variance. Each run recomputed the VarPro fit at the supplied variance. Relative to the correct-variance run, mean risk rose by 11.00\% and 12.72\% at half variance and by 26.39\% and 9.84\% at twice the variance, in four- and two-mode order. This intervention changes the ridge penalty, network features and update scales together.

\section{End-to-End Computational Cost}\label{sec:runtime}
\begin{resulttable}\centering\small
\caption{End-to-end CPU profile for VarPro followed by full refinement (two threads, 24 inputs, three fresh processes). Component means sum to total time. Ranges are across the median repetition for each input.}\label{tab:profile}
\begin{tabular}{@{}lrr@{}}\toprule\rowcolor{TableHeader}
Component & Mean (ms) & Input range (ms)\\\midrule
VarPro initialization & 393.39 & [376.92, 485.45]\\
Input / descriptor & 1.73 & [1.60, 2.37]\\
Tangent construction & 7.52 & [6.71, 9.63]\\
Physical features & 10.17 & [8.57, 12.30]\\
MLP (three stages) & 0.46 & [0.43, 0.55]\\
Correction bases & 5.32 & [4.42, 5.99]\\
LM systems & 9.02 & [8.30, 10.31]\\
Rendering / acceptance & 11.23 & [10.65, 13.70]\\
Other / conversion & 3.26 & [3.13, 3.96]\\
\midrule Total (instrumented) & 442.09 & [426.41, 543.54]\\
\bottomrule\end{tabular}\end{resulttable}

Table~\ref{tab:profile} gives the component profile. We profiled the first four test channels under both masks and all three SNRs (24 inputs). Each of three fresh processes used two Xeon Platinum 8352M threads, batch size one, checkpoint 20261102 and three warmups. Every measured call rebuilt the dictionary and recomputed the VarPro fit. Timing began before input conversion and ended when the final contiguous output array was available. It included all candidate channel evaluations and excluded process startup, model loading and target scoring.

The initialization time includes the path solver and state extraction. The remaining categories cover the Jacobian, features, MLP, correction bases, LM solve and candidate tests. Initial and final channel evaluations, conversion and dispatch are grouped as other costs. Category means sum to mean total time. The instrumented median of the input-level repetition medians was 432.04~ms, compared with 432.42~ms without instrumentation on the same inputs.

The single-thread comparison reported in Section~\ref{sec:noise} uses twelve shared inputs and the same fresh-initialization requirement. The thread counts, sample sizes and aggregation differ between that comparison and the component profile here.

\section{Additional Development Analyses}
\subsection{Choice of initializer}
These exploratory results predate the confirmation study. In a two-fit development comparison at 20~dB, the ratio of mean error for NOMP plus a learned update to that for native VarPro was 3.3535 with four modes and 3.6691 with two. Applying the same fitted update after VarPro on 500 development channels instead gave 0-dB full/VarPro ratios of 0.87225 and 0.86245. The 20-dB intervals included one. These observations motivated the separate 1152-channel test.

On the 500 development channels across seven SNRs, we evaluated 13 estimators per view (91000 predictions). At 0~dB the ratios of full-controller mean error to that of a 24-parameter noise/stage controller were 0.90362 and 0.89764. Against a four-parameter scalar gate they were 0.92764 and 0.92291. These controls differ from the full model in both their capacity and their inputs. The main comparison restricts the inputs while retaining the same hidden layers.

\subsection{Comparison with a grid estimator}
At 20~dB with the two-mode mask, the Grid-DIC estimator had 99th-percentile NMSE of 0.002870, compared with 0.011144 for the NOMP-initialized learned update. The grid estimator used a different representation, arithmetic precision and training budget.

Replacing the path atoms by simplified responses changed both analytic and learned estimates. With access to the complete channel, a dense-grid oracle reached an error near $1.13\times10^{-8}$. The observed estimation errors therefore cannot be attributed solely to a grid-resolution floor.

On an A100 GPU, three repetitions on 24 inputs gave approximately 0.43--0.45~s for NOMP plus learned refinement and 0.018--0.020~s for the native Grid-DIC implementation. These timings include fresh NOMP initialization and every refinement attempt in the learned pipeline.

\subsection{Relation to other observation and prior models}
The two-mode acquisition isolates the role of calibration. It supplies no rows in modes 1 and 2. A path inferred from modes 0 and 3 can nevertheless be rendered in both unobserved modes because the response of each mode is known as a function of arrival direction. An independent unknown gain in each unobserved mode would break this link.

Liang et al.~\cite{liang2025} acquire a common transmit pattern before taking grouped measurements. Related work develops efficient cross-pattern extrapolators~\cite{liang2026} and complex-valued networks for grouped acquisition~\cite{liang2024prnet}. Dinis et al.~\cite{dinis2026pixel} exploit electromagnetic pattern correlations to interpolate the response of a pixel antenna and select its configuration. Here the target comprises four full receive-mode matrices and every observed row comes from the same acquisition.

Other learned estimators place the prior at different points in the inverse problem. UTRR unfolds beamspace sparse recovery~\cite{wu2024trimmed}. DDA-Net operates in Doppler, delay and angle for frequency-hopping signals~\cite{ma2026dda}. RC-Flow combines a learned channel prior with a pilot-domain proximal update~\cite{jiang2026rcflow}. In the present estimator, the learned parameters act on individual path coefficients and angles. The noise variance sets the Levenberg--Marquardt (LM) damping. This differs from predicting the damping itself, as in learned bundle adjustment~\cite{tang2019ba}. Related refinement strategies use a residual network after channel estimation for an intelligent reflecting surface~\cite{gu2023irs} or feedback from decoded data in holographic MIMO~\cite{porto2026code}.

\section{Implementation and Numerical Reproducibility}\label{sec:numerics}
\subsection{Input representation and precision}\label{sec:inputs}
For a batch of size $N_b$ with $K_s$ acquired modes, the input consists of the measurement array $y\in\mathbb C^{N_b\times64\times8}$, the row assignment array in $\mathbb Z^{N_b\times64}$, source descriptors in $\mathbb C^{N_b\times K_s\times2\times129}$, query descriptors in $\mathbb C^{N_b\times4\times2\times129}$ and one noise variance per channel. Observations/descriptors use complex64, owners use integer indices, and the supplied variance uses float32. Descriptor decoding and the numerical path update use double precision. A batch shares its query descriptor table. The VarPro initializer identifies the registered analytic pattern family from the descriptors. Inputs outside that family are rejected, which limits deployment with previously unseen calibration functions.

\subsection{Quadratic physical controls}\label{sec:quadratic}
The coefficient-ridge control replaces the LM coefficient increment with $c_{\rm ridge}-c$, where
\begin{equation}
 c_{\rm ridge}=[A^HA+(\lambda+10^{-10})I]^{-1}A^Hy,
 \qquad \lambda=\kappa L\sigma^2/P_0.
\end{equation}
It retains the LM angular proposal and uses the same channel-increment cap at each of three stages. The joint-quadratic control adds a coefficient penalty to the joint LM system. Define $P_c=\operatorname{diag}(1,1,0,0,\ldots,1,1,0,0)$ and $c_{\mathbb R}=P_cx$. It solves
\begin{equation}
 [G+\mu I+\lambda D_c^{-1}P_cD_c^{-1}]z
 =b-\lambda D_c^{-1}c_{\mathbb R},\qquad
 \delta=D_c^{-1}z.
\end{equation}
The off-diagonal Gram entries couple the path updates in both controls. The grid $\kappa\in\{0,0.1,0.3,1,3,10\}$ selected $\kappa=3$ for each control on the first 100 development channels, pooled over both masks at 0~dB. Native VarPro had the lowest mean risk among these physical controls. The adaptive-ridge comparison in Section~\ref{sec:shrinkage} updates its diagonal penalty from the current coefficients while holding all angles fixed.

The VarPro multiplier was selected as $0.05$ from $\{0,0.005,0.015,0.05,0.15,0.5\}$ on the same development subset. Its objective averages the measurement residual over $m$, whereas the physical controls above use an unaveraged sum. The resulting $m$ factor must therefore be retained when comparing their normal equations. The trust-region-reflective solver uses the analytic reduced-residual Jacobian, \texttt{x\_scale=jac} and $\mathrm{ftol}=\mathrm{xtol}=\mathrm{gtol}=10^{-8}$. Intermediate joint fits allow at most five function evaluations and the final fit allows forty.

\subsection{Scene generation and repeated training}
Regenerating the 5000 training targets, 500 development targets and all 33000 acquisition views reproduced their original arrays bit for bit. No full-channel hash occurred in more than one of the training, development and 1152-channel test sets. On 48 training views, regenerated NOMP states agreed with the original estimates to within $1.2\times10^{-14}$.

A repeated training run for seed 20261102 used all 3750 training minibatches from the original NOMP estimates. The final CPU-trained weights differed from the GPU-trained weights by at most $3.1\times10^{-13}$. The largest absolute difference between training losses over the run was $6.1\times10^{-15}$.

\subsection{Transformation equivariance}
We also tested the three-stage update under path permutations and global phase rotations on two input examples for all five trained fits. Across 40 runs, exact quarter-turn rotations, permutations and their composition agreed with the transformed reference outputs to within $4.1\times10^{-14}$ per entry. A 0.71-rad rotation followed by conversion to the original complex64 input precision gave a discrepancy of at most $3.0\times10^{-8}$. Each run transformed the initial path state consistently with the input.

\subsection{Cross-platform inference checks}
Recomputing the statistical summaries from the archived per-channel errors reproduced the reported results. Fresh inference on the two supplied examples agreed entrywise to within $4.09\times10^{-14}$ on a second CPU platform. A check across the first channel of each acquisition and SNR condition from the confirmation and noise-sweep panels gave a maximum absolute NMSE difference of $5.27\times10^{-9}$ across 170 scores. Sixty-one scores exceeded the original cross-run tolerance (relative $10^{-10}$ and absolute $10^{-12}$). In the full replay for~\eqref{eq:unobserved_nmse}, the largest absolute discrepancy from the archived full-energy regional contributions was $2.23\times10^{-9}$. Adaptive-ridge full-channel NMSE agreed to within $4.05\times10^{-15}$. Training and timing results reported elsewhere in this supplement come from the original experiments.

\bibliographystyle{IEEEtran}\bibliography{references}